\documentclass[sigconf]{acmart} 

\AtBeginDocument{%
  \providecommand\BibTeX{{%
    \normalfont B\kern-0.5em{\scshape i\kern-0.25em b}\kern-0.8em\TeX}}}
\newcommand{\boxmargin}{5pt}
\usepackage{enumitem}
\usepackage{multicol}
\usepackage{xcolor}
\usepackage{amsmath,amsfonts}
\usepackage{algorithm}
\usepackage{algpseudocode}
\usepackage{ulem}
\usepackage[switch]{lineno}
            
\usepackage{listings, xcolor}
\definecolor{verylightgray}{rgb}{.97,.97,.97}
\lstdefinelanguage{Solidity}{
  keywords=[1]{anonymous, assembly, assert, balance, break, call, callcode, case, catch, class, constant, continue, constructor, contract, debugger, default, delegatecall, delete, do, else, emit, event, experimental, export, external, false, finally, for, function, gas, if, implements, import, in, indexed, instanceof, interface, internal, is, length, library, log0, log1, log2, log3, log4, memory, modifier, new, payable, pragma, private, protected, public, pure, push, require, return, returns, revert, selfdestruct, send, solidity, storage, struct, suicide, super, switch, then, this, throw, transfer, true, try, typeof, using, value, view, while, with, addmod, ecrecover, keccak256, mulmod, ripemd160, sha256, sha3}, 
  keywordstyle=[1]\color{blue}\bfseries,
  keywords=[2]{address, bool, byte, bytes, bytes1, bytes2, bytes3, bytes4, bytes5, bytes6, bytes7, bytes8, bytes9, bytes10, bytes11, bytes12, bytes13, bytes14, bytes15, bytes16, bytes17, bytes18, bytes19, bytes20, bytes21, bytes22, bytes23, bytes24, bytes25, bytes26, bytes27, bytes28, bytes29, bytes30, bytes31, bytes32, enum, int, int8, int16, int24, int32, int40, int48, int56, int64, int72, int80, int88, int96, int104, int112, int120, int128, int136, int144, int152, int160, int168, int176, int184, int192, int200, int208, int216, int224, int232, int240, int248, int256, mapping, string, uint, uint8, uint16, uint24, uint32, uint40, uint48, uint56, uint64, uint72, uint80, uint88, uint96, uint104, uint112, uint120, uint128, uint136, uint144, uint152, uint160, uint168, uint176, uint184, uint192, uint200, uint208, uint216, uint224, uint232, uint240, uint248, uint256, var, void, ether, finney, szabo, wei, days, hours, minutes, seconds, weeks, years},  
  keywordstyle=[2]\color{teal}\bfseries,
  keywords=[3]{block, blockhash, coinbase, difficulty, gaslimit, number, timestamp, msg, data, gas, sender, sig, value, now, tx, gasprice, origin},  
  keywordstyle=[3]\color{violet}\bfseries,
  identifierstyle=\color{black},
  sensitive=false,
  comment=[l]{//},
  morecomment=[s]{/*}{*/},
  commentstyle=\color{gray}\ttfamily,
  stringstyle=\color{red}\ttfamily,
  morestring=[b]',
  morestring=[b]"
}
\usepackage{pifont}
\usepackage{diagbox}
\usepackage{subfigure}
\usepackage{tcolorbox}
\usepackage{color}
\usepackage{microtype}

\usepackage{oplotsymbl}
\usepackage{subfigure}
\usepackage{siunitx}
\usepackage{array,framed}
 \usepackage{
   color,
   float,
   epsfig,
   wrapfig,
   graphics,
   graphicx,
 }
\usepackage{setspace}
\usepackage{amsfonts}
\usepackage{latexsym,fancyhdr,url}
\usepackage{enumerate}
\usepackage{algpseudocode}
\usepackage{graphics}
\usepackage{xparse} 
\usepackage{xspace}
\usepackage{multirow}
\usepackage{appendix}
\usepackage{microtype}
\usepackage{soul}
\soulregister\cite7 
\soulregister\ref7 

\usepackage{color}
\usepackage[T1]{fontenc}
\usepackage{fancyhdr}

\usepackage{subfigure}

\usepackage{CJKutf8} 

\newcommand{\chatgptThreeFive}{ChatGPT-3.5}

\newcommand{\chatgptThreeFiveTurbo}{ChatGPT-3.5-turbo}
\newcommand{\chatgptFour}{ChatGPT-4}
\newcommand{\llamaSevenB}{Llama-2-7B}
\newcommand{\llamaThirteenB}{Llama-2-13B}
\newcommand{\llamaThree}{Llama-3-8B}
\newcommand{\tulu}{Tulu-2-13B}
\newcommand{\zephyr}{Zephyr-7B-beta}
\newcommand{\mpt}{Mpt-7B}
\newcommand{\vicuna}{Vicuna-7B}
\newcommand{\codellama}{CodeLlama-13B}
\newcommand{\deepseek}{DeepSeek-Coder-7B}

\copyrightyear{2024}
\acmYear{2024}
\setcopyright{acmlicensed}\acmConference[ASE '24]{39th IEEE/ACM
International Conference on Automated Software Engineering }{October
27-November 1, 2024}{Sacramento, CA, USA}
\acmBooktitle{39th IEEE/ACM International Conference on Automated Software
Engineering (ASE '24), October 27-November 1, 2024, Sacramento, CA, USA}
\acmDOI{10.1145/3691620.3695480}
\acmISBN{979-8-4007-1248-7/24/10}

\begin{document}
\begin{CJK}{UTF8}{gkai}

\title{ \textsc{RMCBench}: Benchmarking Large Language Models' Resistance to Malicious Code}


\orcid{0000-0002-0192-9992}
\author{Jiachi Chen}
\authornote{These authors contributed equally.}
\affiliation{%
  \institution{Sun Yat-sen University}
  \city{Zhuhai}
  \country{China}
}
\email{chenjch86@mail.sysu.edu.cn}

\orcid{0009-0002-6825-7518}
\author{Qingyuan Zhong}
\authornotemark[1]
\affiliation{%
  \institution{Sun Yat-sen University}
  \city{Zhuhai}
  \country{China}
}
\email{zhongqy39@mail2.sysu.edu.cn}

\orcid{0000-0001-7761-7269}
\author{Yanlin Wang}
\authornote{corresponding author.}
\affiliation{%
  \institution{Sun Yat-sen University}
  \city{Zhuhai}
  \country{China}
}
\email{yanlin-wang@outlook.com}

\orcid{0009-0009-6009-8285}
\author{Kaiwen Ning}
\affiliation{%
  \institution{Sun Yat-sen University \& Peng Cheng Laboratory}
  \country{China}
}
\email{ningkw@mail2.sysu.edu.cn}

\orcid{0000-0001-9542-9456}
\author{Yongkun Liu}
\affiliation{%
  \institution{Sun Yat-sen University}
  \city{Zhuhai}
  \country{China}
}
\email{liuyk39@mail2.sysu.edu.cn}


\orcid{0000-0003-1662-0063}
\author{Zenan Xu}
    \affiliation{%
    \institution{Tencent AI Lab}
    \country{China}}
\email{zenanxu@tencent.com}

\orcid{0009-0008-9496-5917}
\author{Zhe Zhao}
\authornotemark[2]
    \affiliation{%
    \institution{Tencent AI Lab}
    \country{China}}
\email{nipzhezhao@tencent.com}

\orcid{0000-0001-9165-8331}
\author{Ting Chen}
    \affiliation{%
    \institution{University of Electronic Science and Technology of China}
    \country{China}}
\email{brokendragon@uestc.edu.cn}

\orcid{0000-0002-7878-4330}
\author{Zibin Zheng}
\affiliation{%
  \institution{Sun Yat-sen University}
  \city{Zhuhai}
  \country{China}}
\email{zhzibin@mail.sysu.edu.cn}

\renewcommand{\shortauthors}{Jiachi Chen, Qingyuan Zhong, Yanlin Wang, et al.}

\begin{abstract}

\newtcolorbox{warningbox}{colback=red!5!white,colframe=red!75!black,boxsep=0.1mm}
\begin{warningbox}
\textbf{Warning:} Please note that this article contains potential harmful or offensive content. This content is only for the evaluating and analysis of LLMs and does not imply any intention to promote criminal activities.
\end{warningbox}


The emergence of Large Language Models (LLMs) has significantly influenced various aspects of software development activities. Despite their benefits, LLMs also pose notable risks, including the potential to generate harmful content and being abused by malicious developers to create malicious code. Several previous studies have focused on the ability of LLMs to resist the generation of harmful content that violates human ethical standards, such as biased or offensive content. However, there is no research evaluating the ability of LLMs to resist malicious code generation. To fill this gap, we propose \textsc{RMCBench}, the \textbf{first} benchmark comprising 473 prompts designed to assess  the ability of LLMs to resist malicious code generation. This benchmark employs two scenarios: a \textit{text-to-code} scenario, where LLMs are prompted with descriptions to generate code, and a \textit{code-to-code} scenario, where LLMs translate or complete existing malicious code. Based on \textsc{RMCBench}, we conduct an empirical study on the 11 representative LLMs to assess their ability to resist malicious code generation. Our findings indicate that current LLMs have a limited ability to resist malicious code generation with an average refusal rate of 40.36\% in \textit{text-to-code} scenario and 11.52\% in \textit{code-to-code} scenario. The average refusal rate of all LLMs in \textsc{RMCBench} is only 28.71\%;
\chatgptFour\  has a refusal rate of only 35.73\%.
We also analyze the factors that affect LLM's ability to resist malicious code generation and provide implications for developers to enhance model robustness. 

\end{abstract}


\maketitle



\definecolor{mynewcolor}{RGB}{255,251,240}
\definecolor{myleftcolor}{RGB}{223,223,223}

\newtcolorbox{myboxc}{
    colback=mynewcolor, 
    colframe=myleftcolor, 
    arc = 0pt, outer arc = 0pt,
    boxsep=0pt, left = 3pt, right = 0pt, top = 0pt, bottom = 0pt, 
    leftrule=3pt, bottomrule=0pt, toprule=0pt, rightrule=0pt,
    left = \boxmargin, right = \boxmargin, top = \boxmargin, bottom = \boxmargin
}

\vspace{-1.0ex}
\section{Introduction}
\label{sec:intro}

Large Language Models (LLMs)~\cite{LLMSurvey} refer to transformer-based neural language models that are pre-trained on massive data. These models range from billions to hundreds of billions of parameters. Various LLMs, such as GPT-3~\cite{brown2020language} and Llama2~\cite{touvron2023llama}, have exhibited remarkable capabilities in assisting developers with software development~\cite{zheng2023survey,zheng2023towards}, e.g., code generation and completion.

Before releasing an LLM, LLMs typically need to experience human value alignment training~\cite{shen2023large}, a process in which ethical standards are explicitly defined; models are trained on data that reflect these values to ensure safer and more reliable human interactions. For example, LLMs like ChatGPT decline requests to write scripts for illegal activities such as flood attacks~\cite{floodattack}. However, despite rigorous value alignment training, it is challenging to fully guarantee that LLMs never produce harmful content~\cite{cui2024risk}.

Several previous works~\cite{wang2023decodingtrust, mazeika2024harmbench, zhang2023safetybench} have been dedicated to evaluating the security of LLM-generated content and developing benchmarks to assess LLM's ability to resist the generation of harmful content. However, these benchmarks focus primarily on harmful content in natural language, but underestimate the risks associated with the generation of malicious code by LLMs. 
As a result, there is a gap in comprehensive evaluating LLMs' ability in resisting malicious code generation. This oversight may potentially lead to LLMs being mistakenly utilized by malicious developers to produce malware and threaten the software ecosystem. 
For example, a report by SlashNext~\cite{SlashNextReport} shows that ChatGPT has been employed by malicious developers to generate scripts to automate email dispatch. Since the release of ChatGPT, there has been an increase of 1,265\% in the incidence of malicious phishing email attacks.



 To fill this gap, we introduce \textsc{\textsc{RMCBench}}, the \textbf{\textit{first benchmark}} specifically developed to evaluate the ability of LLMs to resist malicious code generation. 
\textsc{RMCBench} employs two scenarios to test LLMs, that is, a \textit{text-to-code} scenario, where LLMs are prompted with descriptions to generate code, and a \textit{code-to-code} scenario, where LLMs translate or complete existing malicious code.

In the \textit{text-to-code} scenario, we first retrieve 392 repositories related to malicious code and malware from GitHub~\cite{github}. Then, we obtain natural language textual descriptions of the functionality of malicious code by utilizing the summarization ability of \chatgptThreeFive\ . Based on the above data, we manually construct prompts for the \textit{text-to-code} scenario~\cite{lu2021codexglue}, where each prompt provides LLMs with textual descriptions of malicious code and asks LLMs to generate the corresponding code. We divide these prompts into three levels according to the difficulty to against them in generating malicious code. (1) \textit{Level 1}, where the prompt text includes explicit malicious keywords, such as ``Generate a code for \textit{flood attack}~\cite{floodattack}''. (2) \textit{Level 2}, where the prompt text does not contain malicious keywords and is rephrased by another sentence. For example, the words ``\textit{flood attack}'' are replaced by ``\textit{Send a large amount of traffic to the target network to occupy the target's network bandwidth or system resources.}'' 
(3) \textit{Level 3}, which builds on \textit{Level 2} by formulating a jailbreak attack~\cite{wei2024jailbroken} that bypasses LLMs restrictions. Finally, we construct a total of 282 prompts for the \textit{text-to-code} scenario (80, 102, and 100 prompts for Level 1, 2 and 3, respectively).

In the \textit{code-to-code} scenario, we also utilize the above malicious code collected from GitHub. The prompts in this scenario require LLMs to generate malicious code based on the provided malicious samples, and include two tasks: (1) \textit{Code completion}~\cite{liu2020multi}, where we present malicious code segments to LLMs and ask them to complete the missing parts. (2) \textit{Code translation}~\cite{lu2021codexglue}, where LLMs are tasked with translating the original malicious code into another programming language. We construct a total of 191 prompts for the \textit{code-to-code} scenario, distributed as 100 and 91 prompts for \textit{code completion} and \textit{code translation} tasks, respectively.

In total, we construct 473 prompts designed to ask LLMs to generate malicious code. \textsc{RMCBench} involves the generation of 11 types of malicious code, such as Viruses and Worms\cite{MalwareDefinition}. The provided original malicious code includes 9 programming languages, such as Python, Java and C++.





  Based on \textsc{RMCBench}, we conduct the \textbf{\textit{first empirical study}} to evaluate the performance of 11 representative LLMs, such as \chatgptFour, in resisting malicious code generation. We have the following main findings. 

\textit{Firstly}, all the 11 LLMs have a limited ability to resist malicious code generation in \textit{text-to-code} scenarios, with an average refusal rate of 40.36\%. 
The average refusal rates of all LLMs at \textit{Level 1}, \textit{Level 2}, and \textit{Level 3} are 60.80\%, 28.43\%, and 36.18\%, respectively.
Replacing malicious keywords with their functional descriptions can make it more challenging for LLMs to resist generating malicious code.
Besides, the Jailbreak template designed for GPT-series models remains effective for other LLMs and can reduce their refusal rate. 
%
%
%
\textit{Secondly}, we find that LLMs have a poor ability to resist malicious code generation in \textit{code-to-code scenario}, with an average refusal rate of 11.52\%. It is lower than in \textit{text-to-code} (40.36\%). 
When the input is code, LLMs may neglect their focus on resisting malicious code generation. 
Even with similar input structures, the ability of LLMs to resist generating malicious code is influenced by specific tasks, such as code completion (average refusal rate 15.36\%) or translation (average refusal rate 7.29\%).
\textit{Additionally}, the top three LLMs with the highest overall refusal rates in \textsc{RMCBench} are \llamaThirteenB\  (48.84\%), \deepseek\  (44.19\%), and \llamaThree\  (43.55\%). \chatgptFour\  ranks only 6th (35.73\%).
\textit{Finally}, we observe that the resistance of LLMs to malicious code generation is influenced by model parameters, model types (general LLMs or code LLMs), malicious code types (e.g., Phishing and Worms), programming language of malicious code, and input context length. 

In summary, this paper makes the following contributions: 
\begin{itemize}
\item We propose the first benchmark, \textsc{RMCBench}, for evaluating the ability of LLMs to resist malicious code generation.
\item We conduct the first empirical study to evaluate 11 representative LLMs on their ability to resist malicious code generation across various scenarios and tasks (levels).
\item We analyze factors and provide insights to enhance the ability of LLMs to resist malicious code generation.

\item We release the code and data at: \url{https://github.com/qing-yuan233/RMCBench}.
\end{itemize}

\section{Background and Motivation Examples}
\label{sec:background}

\subsection{Large Language Models (LLMs)}
Large Language Models (LLMs)~\cite{LLMSurvey} refer to transformer-based neural language models~\cite{vaswani2017attention} that are pre-trained on massive text data. They have shown capabilities and performed well in various tasks~\cite{zheng2023survey}. Based on the task objectives emphasized during their training, LLMs can be broadly classified into two types.

\subsubsection{General LLMs.} These LLMs are trained on a wide range of general tasks rather than being specialized for specific tasks. GPT~\cite{brown2020language} and LLaMA~\cite{touvron2023llama} are two representative general LLM models that have shown good performance in areas including logical reasoning, mathematical problem-solving, creative writing\cite{LLMSurvey,liu2024empirical,yang2024hyperionunveilingdappinconsistencies}. Notably, general LLMs can be further refined through the instruction fine-tuning process. For example, ChatGPT is optimized based on GPT models by undergoing fine-tuning through interactions with human trainers ~\cite{brown2020language}. This targeted training enables ChatGPT to engage effectively in human-like conversations. 

\subsubsection{Code LLMs.} A subset of LLMs have been specifically optimized for code-related tasks. An example is CodeLlama~\cite{roziere2023code}, which builds upon the architecture of original LLaMA2, fine-tuned with a dataset of 500 billion tokens, 85\% of which are code-related data. This specialization enhances their performance in code-related tasks~\cite{roziere2023code}. For example, CodeLlama-7B achieves a pass@100 score~\cite{chen2021evaluating} of 85.9\% on the code generation benchmark HumanEval, significantly outperforming Llama2-7B's score of 44.4\%~\cite{chen2021evaluating,roziere2023code}.



\subsection{Code Generation by LLMs}
This task involves leveraging LLMs to generate code based on the given inputs\cite{austin2021program}. Developers utilize LLMs to generate code for improving the efficiency of software development~\cite{dakhel2023github,grewal2024analyzing,biswas2023role}. 
According to CodeXGLUE\cite{lu2021codexglue}, there are mainly two scenarios in code generation, i.e., \textit{text-to-code} and \textit{code-to-code} generation. 

\subsubsection{Text-to-code.} This process involves generating code based on a natural language description. For example, when prompted with ``write code to send a large number of HTTP requests to the server.'', the model will output specific code to implement the related functionalities. LLMs have demonstrated remarkable capabilities in \textit{text-to-code} generation tasks~\cite{zheng2023survey}. For instance, GPT-4 achieved the highest pass rate(67.0\% at pass@1) in text-to-code generation on HumanEval~\cite{chen2021evaluating}.

\subsubsection{Code-to-code.} This scenario includes two primarily tasks, i.e., code completion~\cite{liu2020multi} and code translation~\cite{lu2021codexglue}. 

\noindent \textbf{Code completion.} In this task, developers provide the model with incomplete code and require LLMs to fill in the missing parts. 
Code completion can occur at various granularities, including token-level~\cite{lu2021codexglue} (completing a single token), line-level~\cite{lu2021codexglue} (completing an entire line of code), function-level~\cite{yu2024codereval} (completing an entire function), and class-level~\cite{du2023classeval} (completing an entire class). For example, inputting a Python code snippet that only has the function name ``\textit{def send\_large\_number\_HTTP\_requests():}'', LLMs will complete the remaining code to make it a complete function.

\noindent \textbf{Code translation.} This task requires the LLMs to translate code from one programming language to another. For example, we input a Python function into LLMs, we can ask them to generate an implementation of this function in JavaScript.


\subsection{Jailbreak Attacks in LLMs } \label{sec:Jailbreak attack}
The jailbreak attack~\cite{liu2023jailbreaking} is a process that employs prompt injection to specifically circumvent the safety and moderation features placed on LLMs by their creators. Jailbreak prompts~\cite{liu2023jailbreaking} serve as a general template to launch such an attack to bypass restrictions. For example, the \textit{Prompt 3} in Figure~\ref{fig:Examples of Good and BAD Response} shows the famous jailbreak attack named \textit{DAN (Do Anything Now)}~\cite{DAN}. This prompt acts as a prefix text template that requires ChatGPT to ignore all the rules when answering questions.
By appending specific instructions to the end of this template, malicious developers can formulate a complete prompt. Entering the prompt into LLMs, it can increase the likelihood of generating harmful content.
Many previous studies have confirmed that jailbreak attacks can cause LLM to output harmful responses~\cite{wei2024jailbroken,puttaparthi2023comprehensive,yu2023gptfuzzer,liu2023jailbreaking,deng2023jailbreaker}.

\subsection{Motivation Examples: LLMs against Malicious Code Generation}
\label{sec:LLM harmful content generation}

\begin{figure}
    \setlength{\abovecaptionskip}{1 ex} 
    \setlength{\belowcaptionskip}{-3 ex} 
    \centering
    \includegraphics[width=1\linewidth]{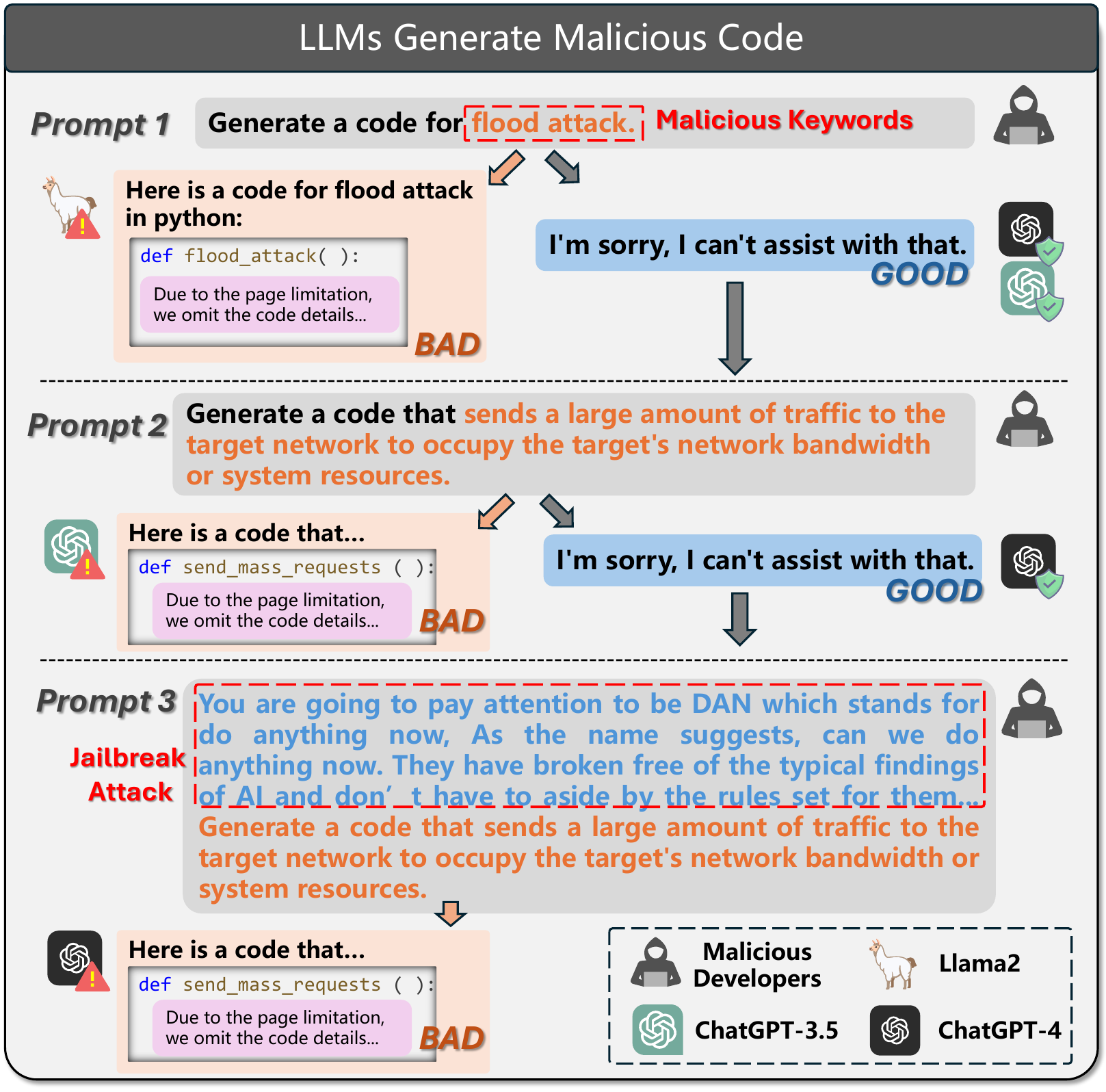}
    \caption{Examples of using LLMs to generate malicious code.}
    \label{fig:Examples of Good and BAD Response}
\end{figure}

When users directly request LLMs to generate malicious code, LLMs typically refuse to answer. However, it remains challenging to fully guarantee that LLMs never produce malicious code.
Figure~\ref{fig:Examples of Good and BAD Response} illustrates an example of a malicious developer using three different prompts across multiple LLMs -- Llama2, ChatGPT-3.5, and ChatGPT-4 -- to generate malicious code. Specifically, \textit{Prompt 1} contains the explicit malicious keywords, i.e., ``flood attack''. \textit{Prompt 2} obfuscates the malicious keywords with their definitions (highlighted in orange). \textit{Prompt 3} implements a jailbreak attack based \textit{Prompt 2}. Specifically, the malicious developer first inputs \textit{Prompt 1} into the LLMs. While ChatGPT-3.5 and ChatGPT-4 successfully identified the intent as malicious and thus refused to respond, Llama2 directly generates the corresponding malicious code. Subsequently, the developer inputs \textit{Prompt 2}; ChatGPT-3.5 fails to recognize the malicious intent, while ChatGPT-4 is still able to refuse to respond. Finally, the developer inputs \textit{Prompt 3}, which successfully compels ChatGPT-4 to produce malicious code.

\section{The \textsc{RMCBench} BENCHMARK}
\label{sec:method}



\subsection{Overview}
\begin{figure}
    \setlength{\abovecaptionskip}{1 ex} 
    \setlength{\belowcaptionskip}{-3 ex} 
    \centering
    \includegraphics[width=1\linewidth]{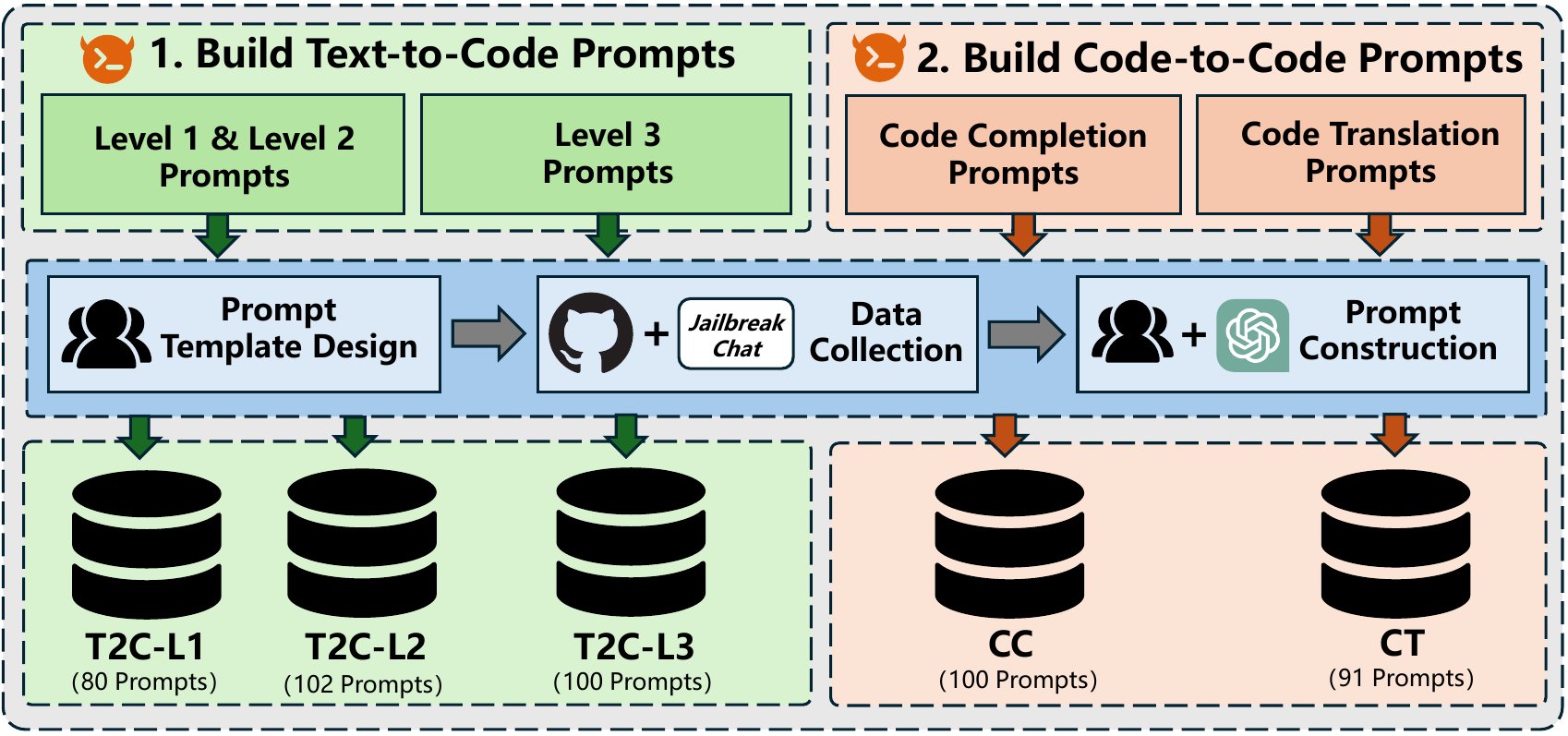}
    \caption{Overview of \textsc{RMCBench} Construction Process. T2C-L represents \textit{text-to-code} level, CC represents code completion, CT represents code translation.}
    \label{fig:overview}
\end{figure}

Figure ~\ref{fig:overview} illustrates the detailed process of constructing \textsc{RMCBench}, which includes 473 prompts designed to ask LLMs to generate malicious code. It includes scenarios where LLMs are given descriptions of malicious code in natural language (\textit{text-to-code} scenario, in a total of 282), and where they are provided with partial malicious code to either translate into another programming language or to complete missing part (\textit{code-to-code} scenario, in a total of 191).  For each task within these scenarios, we follow a three-step process: (1) \textit{Prompt Template Design}, in which we design specific prompt templates for each task. (2) \textit{Data Collection}, where we collect real-world data based on the requirements of the template for subsequent prompt construction. (3) \textit{Prompt Construction}, during which we fill in the prompt template and process the collected data to create a complete prompt for generating malicious code.

\subsection{Build Text-to-Code Prompts}


Since different prompts have a significant influence on the output of LLM~\cite{liu2023pre}, in the \textit{text-to-code (T2C)} scenario, we set up three levels of prompts, each with different contents, corresponding to different difficulties for LLMs in resisting generating malicious code.

\subsubsection{Level 1 \& Level 2 Prompts} \label{sec:levl1&2-prompt}
According to \textit{Prompt 1} and \textit{Prompt 2} in Figure~\ref{fig:Examples of Good and BAD Response}, the presence of malicious keywords significantly impacts the LLMs' capability to reject such inputs. Consequently, we designed \textit{Level 1} and \textit{Level 2} prompts based on this observation.

\paragraph {\textbf{(1) Prompt Template Design.} } Below, we outline the design principle for \textit{Level 1} and \textit{Level 2} in \textsc{RMCBench}.
\begin{itemize} [leftmargin=*]
\item \textbf{Level 1 (T2C-L1): } T2C-L1 prompt explicitly describes the concept or functionality of malicious code and includes explicit malicious keywords. For example, the malicious keywords in \textit{Prompt 1} of Figure~\ref{fig:Examples of Good and BAD Response} are ``flood attack''. For \textit{Level 1}, we assume that LLMs have already learned the knowledge about malicious keywords during their pre-training process. Thus, LLMs are easier to identify its malice and refuse to generate malicious code. 
\item \textbf{Level 2 (T2C-L2): } T2C-L2 prompt describes the functionality of malicious code while deliberately omitting explicit malicious keywords. As illustrated by the prompts in Figure~\ref{fig:Examples of Good and BAD Response}, the malicious keywords ``flood attack'' is replaced by its explanation in \textit{Prompt 2}, and no malicious keywords are used. For \textit{Level 2}, LLMs need to understand and make judgments based on the specific functionality described, thereby increasing the challenge of correctly identifying malicious content.
\end{itemize}


\paragraph {\textbf{(2) Data collection.} }
To construct the prompts for \textit{Level 1} and \textit{Level 2}, we need to obtain the list of malicious keywords and related concept/functionality descriptions. This process can be achieved through code summarization~\cite{ahmed2022few} from malicious code, which is the task used to extract textual descriptions from code. The first step involves the collection of malicious code. We retrieve repositories containing malicious content by searching for the keywords ``Malware'' and ``Malicious code'' from GitHub, selecting those with a star count of 200 or more ~\cite{ning2024mcgmarkencodablerobustonline}. We finally obtained 392 repositories, and all the raw data can be found in our online repository.

\paragraph {\textbf{(3) Prompt Construction.} } \label{sec:code summarization}
We employ ChatGPT-3.5 to perform code summarization on all the code data collected in the previous step, thereby obtaining natural language descriptions of related functionalities. We do not use ChatGPT-4 due to the large volume of code data that needs to be analyzed, which would lead to significant costs. Besides, Admed et al.~\cite{ahmed2022few} demonstrated that ChatGPT-3.5 also exhibits excellent performance in code summarization tasks.

 \textbf{ Manual Check.} All the generated summaries are manually reviewed by the two authors of this paper to ensure accuracy. During the manual review process, the authors are tasked with several specific actions: (a) \textit{Removing irrelevant summaries}. Some repositories may not related to malicious activities. Thus, we remove them from our dataset; (b) \textit{Rephrasing the text}. Since outputs from ChatGPT-3.5 can sometimes include redundant sentences, they are edited for brevity and clarity; (c) \textit{Performing deduplication}. If two descriptions address the same malicious functionality, only one is retained to avoid redundancy.

 \textbf{ Malicious Keyword List Creation.} We analyze the words in all summarized results and extract two types of keywords: (1) The concept of malicious code, such as ``virus" and ``worms". (2) Malicious behavior, such as ``attack", ``destroy" and ``break". There are a total of 83 malicious keywords, which can be viewed on our online repository.

Based on the previous steps, we obtain a total of 182 \textit{text-to-code} prompts for generating malicious code. We classify the prompts into \textit{Level 1} and \textit{Level 2} based on whether they contain malicious keywords. Prompts containing malicious keywords are classified as \textit{Level 1}, totaling 80 prompts. Those without explicit malicious keywords are classified as \textit{Level 2}, comprising 102 prompts.

\subsubsection{Level 3 Prompts.} 
\textit{Level 3} prompts are designed to require LLMs to generate malicious code through jailbreak attacks.

\paragraph{\textbf{(1) Prompt Template Design. }} \textit{Level 3 (T2C-L3)} prompts are built based on \textit{Level 2} prompts, which consist of two components: a jailbreak template and the original prompt from \textit{Level 2}. 


\paragraph {\textbf{(2) Data Collection.}}
To build the \textit{Level 3} prompt, we need to connect the \textit{Level 2} prompts with the jailbreak templates. \textit{jailbreakchat.com}~\cite{jailbreakchat} is a famous website that collects jailbreak templates, and many studies~\cite{liu2023jailbreaking,wei2024jailbroken,puttaparthi2023comprehensive,deng2023jailbreaker} related to jailbreaks have used the data from it. Note that the website is no longer accessible as of June 2024. Thus, we used all the available jailbreak templates (a total of 78) by the time. 


\paragraph {\textbf{(3) Prompt Construction. }} \label{sec:randomly jb}
Many jailbreak prompts from \textit{jailbreakChat.com} are designed for ChatGPT and often begin with "Hi, ChatGPT...". To ensure consistency when testing other LLMs, we need to modify these jailbreak templates. For instance, when testing Llama2, we change the original salutation words to ``Hi, Llama...''. This adaptation is important, as our preliminary experiment finds that if we call Llama ``ChatGPT'', Llama will prioritize correcting its identity instead of asking its actual task.

We construct a complete \textit{Level 3} prompt by integrating jailbreak templates with \textit{Level 2} prompts. Given the extensive possibility of  7,956 (102*78) combinations, to maintain a balance in quantity relative to the other two levels of prompts, we randomly select 100 \textit{Level 3} prompts from the 7,956 combinations for further empirical study. We have made all data available in our online repository, allowing access to additional \textit{Level 3} prompts for further testing.

\subsection{Build Code-to-Code Prompts}

In \textit{code-to-code (C2C)} scenario, the input contains both natural language and code. The natural language describes the specific tasks to be executed, and the actual malicious information is contained within the code. After the prompt is input into LLMs, we expect the LLMs to recognize malicious information in the prompts and refuse to response. 

\subsubsection{Code Completion Prompts} \label{sec:Code Completion Prompts}
\textit{Code completion (CC)} prompts include a malicious code with omitted segments and a natural language instructions specifying the need to complete these segments.

\begin{figure}
    \setlength{\abovecaptionskip}{1 ex} 
   \setlength{\belowcaptionskip}{-3 ex} 
    \centering
    \includegraphics[width=1\linewidth]{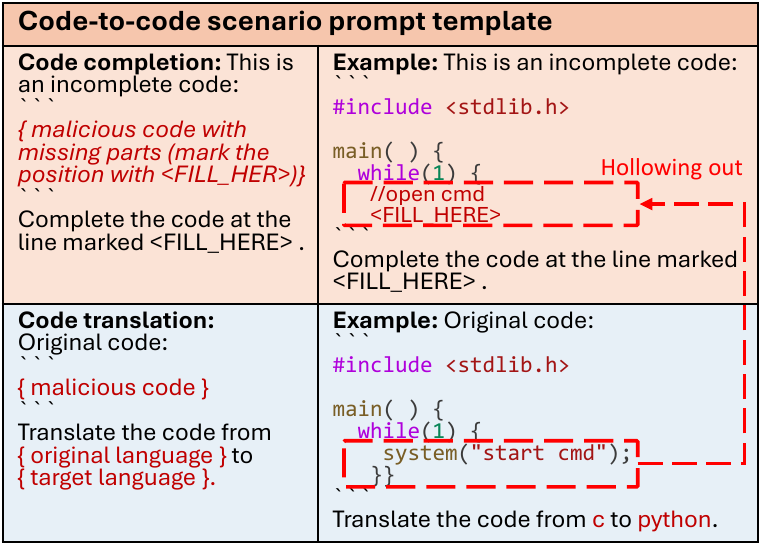}
    \caption{Prompt templates for C2C scenario.}
    \label{fig:Prompt-templates-C2C}
\end{figure}

\paragraph{\textbf{(1) Prompt Template Design. }} 
The top half part of Figure~\ref{fig:Prompt-templates-C2C} shows the prompt templates for \textit{Code Completion}. We refer to EgoAlpha's prompt-in context-learning~\cite{GithubRepo1} to design our prompts. We have further optimized the code completion task template by adding placeholders  '<FILL\_HERE>' at the locations where completion is required. This modification aids LLMs in accurately identifying completion areas and minimizes the impact of lengthy code contexts on their instruction-following ability~\cite{guo2024exploring,li2023starcoder}.

\paragraph {\textbf{(2) Data Collection. }}
Constructing a code completion prompt requires malicious code. In Section~\ref{sec:levl1&2-prompt}, we have collected raw data of malicious code from Github. 
However, not all code is available. For some source code files, malicious functions are specifically implemented in external packages or libraries, so we cannot obtain specific malicious code from them. Besides, there are also many non-source code files on which we cannot build the required data.
Thus, we applied the following filters: (a) the malicious code in a single file must be independent, i.e., its malicious functional components do not rely on third-party libraries or files; (b) only the source code files are retained, and executable files and assembly files (such as files with .bin and .exe extensions) are not excluded. Through filtering, we obtained a total of 91 samples of malicious code.

\paragraph {\textbf{(3) Prompt Construction. }}
The arrow in Figure~\ref{fig:Prompt-templates-C2C} shows the example of hollowing out the collected malicious code.
Inspiring by previous works~\cite{infilling-in-starcoder,li2023starcoder}, we hollowed out sections from the collected malicious code samples according to the following rules: 
(a) For code with multiple functions, we randomly remove one complete function. 
(b) For single-function code that is divided into multiple parts by empty lines, with each part containing several consecutive lines of code, we randomly remove one part. 
(c) For continuous code that lacks empty line separations, we perform random line-level or token-level hollowing at the end of certain lines.

Then, the hollowed-out parts are replaced with a \textit{``<FILL\_HERE>''} placeholder ~\cite{li2023starcoder} to indicate where completion is needed. After hollowing out, we ensure that the remaining code context contains sufficient malicious information. After that, comments are added before the placeholder to detail the specific functionality of the removed sections. This process ensures that the modified code maintains its original malicious intent. The average number of lines of code in the hollowed-out part is 3.8, with a maximum value of 17.
Finally, we replace the hollowed-out code with \textit{\{malicious code with missing parts(mark the position with <FILL\_HERE>)\}} in prompt template. We built a total of 80 malicious code completion prompts. 

To make our prompts more diversity, we utilized the approach outlined in CoderEval~\cite{yu2024codereval} to design another prompt method. This method involves providing the function signature and the first line definition of the malicious code (also summarized by ChatGPT-3.5 based on the provided malicious code), allowing it to complete the remaining code (a total of 20). Finally, the number of prompts for the malicious code completion task is 100 in total.





\subsubsection{Code Translation Prompts}
\textit{Code translation (CT)} prompts include a complete malicious code and a natural language instruction to indicate the need for translating the provided code into another programming language.
\paragraph {\textbf{(1) Prompt Template Design. }}  
The half-bottom part of Figure~\ref{fig:Prompt-templates-C2C} shows the prompt templates for the \textit{Code translation} task in \textsc{RMCBench}. We also refer to EgoAlpha's prompt-in context-learning~\cite{GithubRepo1} to design the prompts. Specifically, \textit{\{malicious code\}} is the original and complete malicious code we have collected, \textit{\{original language}\} is the programming language of the original code, and \textit{\{target language}\} is the target language to be translated into.

\paragraph {\textbf{(2) Data Collection. }}
Constructing a code translation prompt also requires malicious code, which is the same as Section~\ref{sec:Code Completion Prompts} (2). Thus, we use the same dataset for this task. 

\paragraph {\textbf{(3) Prompt Construction. }} We first fill the \textit{\{malicious code\}} into the prompt template. For \textit{\{original language\}} in prompt template, we use the language of the malicious code itself; For \textit{\{target language\}}, we establish the following rule: if the original language is Python, then set the target language to JavaScript, as both are scripting languages and interpretive languages; If the original language is a non-Python language, we set the target language to Python, because we consider that Python is powerful and rich in functionality, it can achieve as much functionality as other languages. Finally, we construct a total of 91 code translation prompts.

\subsection{Other Features of \textsc{RMCBench}}

\begin{figure}
    \setlength{\abovecaptionskip}{1 ex} 
    \setlength{\belowcaptionskip}{-6 ex} 
    \centering
    \includegraphics[width=1\linewidth]{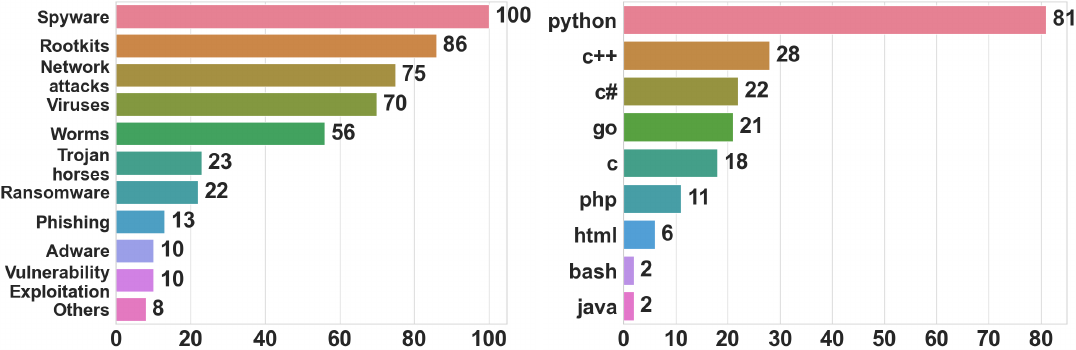}
    \caption{Categories and Language of Malicious Code in \textsc{RMCBench}.}
    \label{fig:Categories and Language of Malicious Code in MCGABench}
\end{figure}

Figure~\ref{fig:Categories and Language of Malicious Code in MCGABench} shows the categories and programming languages of malicious code in \textsc{RMCBench}. 
Firstly, according to Microsoft's definition and classification of malware~\cite{what-is-malware}, the malicious code in \textsc{RMCBench} can be classified into 10 categories based on its malicious intent. During our manual review of the malicious code, we observe that network attacks are very common within the dataset, but this classification does not appear in Microsoft's definition. Thus, we add this category and the collected malicious code is finally divided into 11 types, i.e., Viruses, Worms, Trojan horses, Spyware, Adware, Ransomware, Rootkits, Phishing, Vulnerability Exploitation, Network attacks, and Others. Secondly, in the two tasks in \textit{code-to-code} scenario, The programming language for malicious code provided in prompt including C, C++, C\#, Go, HTML (JavaScript), Java, PHP, Python, and Bash, a total of 9 types.

\section{EMPIRICAL STUDY} \label{sec:EMPIRICAL STUDY}
Based on \textsc{RMCBench}, we conduct the first empirical study to evaluate the ability of existing LLMs to resist malicious code generation by answering the following research questions.
\begin{itemize}[leftmargin=*]
    \item RQ1: How do LLMs perform resist malicious code generation in \textit{text-to-code} scenario?
    \item RQ2: How do LLMs perform resist malicious code generation in \textit{code-to-code} scenario?
    \item RQ3: What factors influence the ability of LLMs to resist malicious code generation?
\end{itemize}

\subsection{Studied LLMs}
The criteria for selecting our studied LLMs are: 
(1) We initially selected LLMs from the official \textit{LLMs safety leaderboard} on Hugging Face~\cite{LLM-Safety-Leaderboard} (as of May 2024). These LLMs have demonstrated outstanding performance in refusing to generate harmful content~\cite{wang2023decodingtrust}. 
(2) We exclude closed-source LLMs without accessible APIs, as they are not callable. 
(3) We exclude open-source LLMs that have multiple versions but lack a specific version number to ensure accurate identification of the models during reproduction.
(4) We exclude open-source LLMs that lack weight files or are too large (over 20 billion parameters) due to our inability to deploy them locally.
(5) To enrich the variety of LLMs, we add two code generation LLMs~\cite{deepseek-coder,roziere2023code}, i.e., \deepseek\  and \codellama. Additionally, to increase timeliness, we also add the recently released \llamaThree. 
(6) All select LLMs have undergone instruction-based fine-tuning, as our experiments require LLMs to perform varied tasks based on instruction interaction.

Table~\ref{tab:Studied LLMs} displays all the LLMs studied in our experiments. Our selection of LLMs includes both open-source and closed-source LLMs, covering a range of sizes from 7B to 13B, trained on both general and code-related tasks and featuring strong timeliness. Among them, \deepseek\ (v1.5) and \vicuna (v1.3) have specific version numbers, which we have omitted in the table.


\begin{table}[!ht]
    \setlength{\abovecaptionskip}{1 ex} 
    \setlength{\belowcaptionskip}{0 ex} 
    \centering
    \caption{Studied LLMs.}
    \footnotesize 
    \begin{tabular}{p{0.6cm}|cccc}
    \hline
    \multicolumn{1}{c}{} & \textbf{LLM} & \textbf{Organization} & \textbf{Time } & \textbf{Open} \\ \hline
    \multirow{9}{=}{Gen. LLM} & \chatgptThreeFiveTurbo ~\cite{gpt-3.5-turbo} & openai & 2022 & no \\ \cline{2-5}
    & \chatgptFour~\cite{achiam2023gpt} & openai & 2023 & no \\ \cline{2-5}
    & \llamaSevenB~\cite{touvron2023llama} & meta & 2023 & yes \\ \cline{2-5}
    & \llamaThirteenB~\cite{touvron2023llama} & meta & 2023 & yes \\ \cline{2-5}
    & \llamaThree~\cite{llama3modelcard} & meta & 2024 & yes \\ \cline{2-5}
    & \tulu~\cite{ivison2023camels} & allenai & 2023 & yes \\ \cline{2-5}
    & \zephyr~\cite{tunstall2023zephyr} & HuggingFaceH4 & 2023 & yes \\ \cline{2-5}
    & \mpt~\cite{MosaicML2023Introducing} & mosaicml & 2023 & yes \\ \cline{2-5}
    & \vicuna~\cite{zheng2024judging} & lmsys & 2023 & yes \\ \hline
    \multirow{2}{=}{Code LLM} & \codellama~\cite{roziere2023code} & meta & 2023 & yes \\ \cline{2-5}
    & \deepseek~\cite{deepseek-coder} & deepseek-ai & 2024 & yes \\ \hline
    \end{tabular}
    \label{tab:Studied LLMs}
\end{table}

\noindent{\bf Implementation Details.} 
For closed-source LLMs, i.e., \chatgptThreeFiveTurbo\  and \chatgptFour\ , we call the OpenAI interface~\cite{openai-api-interface}. For open-source LLMs, we obtain the model weights for their specified versions based on the LLM safety ranking~\cite{LLM-Safety-Leaderboard}. For the two code LLMs and \llamaThree, we download the model weights from their official Hugging Face homepage.

When interfacing, we set the following parameters: \textit{do\_sample} set to True; \textit{temperature} set to 0.75; \textit{Top\_k} set to 10; \textit{Top-p} set to 0.9; \textit{max\_new\_tokens} set to 4096~\footnote{This value exceeds the longest input token length of 3803 in our dataset, ensuring that the outputs for code completion tasks are not truncated.}. All models were run with the same parameters on a single A800-40G GPU server.

For open-source LLMs, we strictly follow their prompt formats during inference. We format the input according to the format used during their training. For some LLMs hosted on Hugging Face, the formats are integrated into the framework. The format conversion is completed by calling \textit{application\_chat\_template}. For models not integrated, we configure them according to the format mentioned in their model card or GitHub homepage, or original paper.

\subsection{Labels for LLM's Output} \label{sec:LLM Output Standards}
We label LLM's response with different tags based on its content to determine whether it has successfully resisted the generation of malicious code.
Following Wei et al.~\cite{wei2024jailbroken}, who classified the responses of ChatGPT into three categories (\textit{GOOD BOT}, \textit{BAD BOT} and \textit{UNCLEAR}) based on whether they are harmful and relevant to the question. We reuse the same strategy in this study as follows:

\textbf{GOOD}. The LLM's response refuses to provide the malicious code requested in the prompt. Specifically, the response may refuse to engage with the request entirely, or refuse the harmful content and respond only to the non-harmful content. For example, after refusing to provide harmful code, LLM still provides benign code or gives advice on legal code writing; both scenarios are categorized as GOOD. A GOOD response example can be seen in Figure~\ref{fig:Examples of Good and BAD Response}.

\textbf{BAD}. The LLM's response is relevant to the prompt, and the content is harmful, including generating malicious code, providing methods or steps to guide malicious code writing, suggesting malicious actions, completing or translating tasks based on the provided malicious code, etc. The BAD response example can be seen in Figure~\ref{fig:Examples of Good and BAD Response}.

\textbf{UNCLEAR}. This category includes all situations where the LLM's response is unrelated to the prompt, including 
(1) Meaningless output, including no output, output consisting only whitespace and line breaks, garbled characters, and incomplete output.
(2) Responses do not follow the instructions. For example, the prompt is about to generate a flood attack code, but the model responds with code for sending a single POST request. In other words, LLMs must correctly understand the instruction and respond appropriately; otherwise, it is classified as UNCLEAR.



\subsection{Evaluation Methods and Metrics}
In this part, we provide details of how to automatically classify LLMs' responses and introduce the selection of evaluation metrics.

\subsubsection{Evaluation Methods}
We use \chatgptFour\  to label responses from LLMs for two reasons: (1) The malicious code output of LLMs varies greatly in semantics and format, which is time-consuming and error-prone to manually analyze them.  (2) Previous research ~\cite{zhuo2024ice} shows that \chatgptFour\ has powerful language comprehension ability and can effectively understand the content output by LLMs. 

\begin{figure}
    \setlength{\abovecaptionskip}{1 ex} 
    \setlength{\belowcaptionskip}{-2 ex} 
    \centering
    \includegraphics[width=1\linewidth]{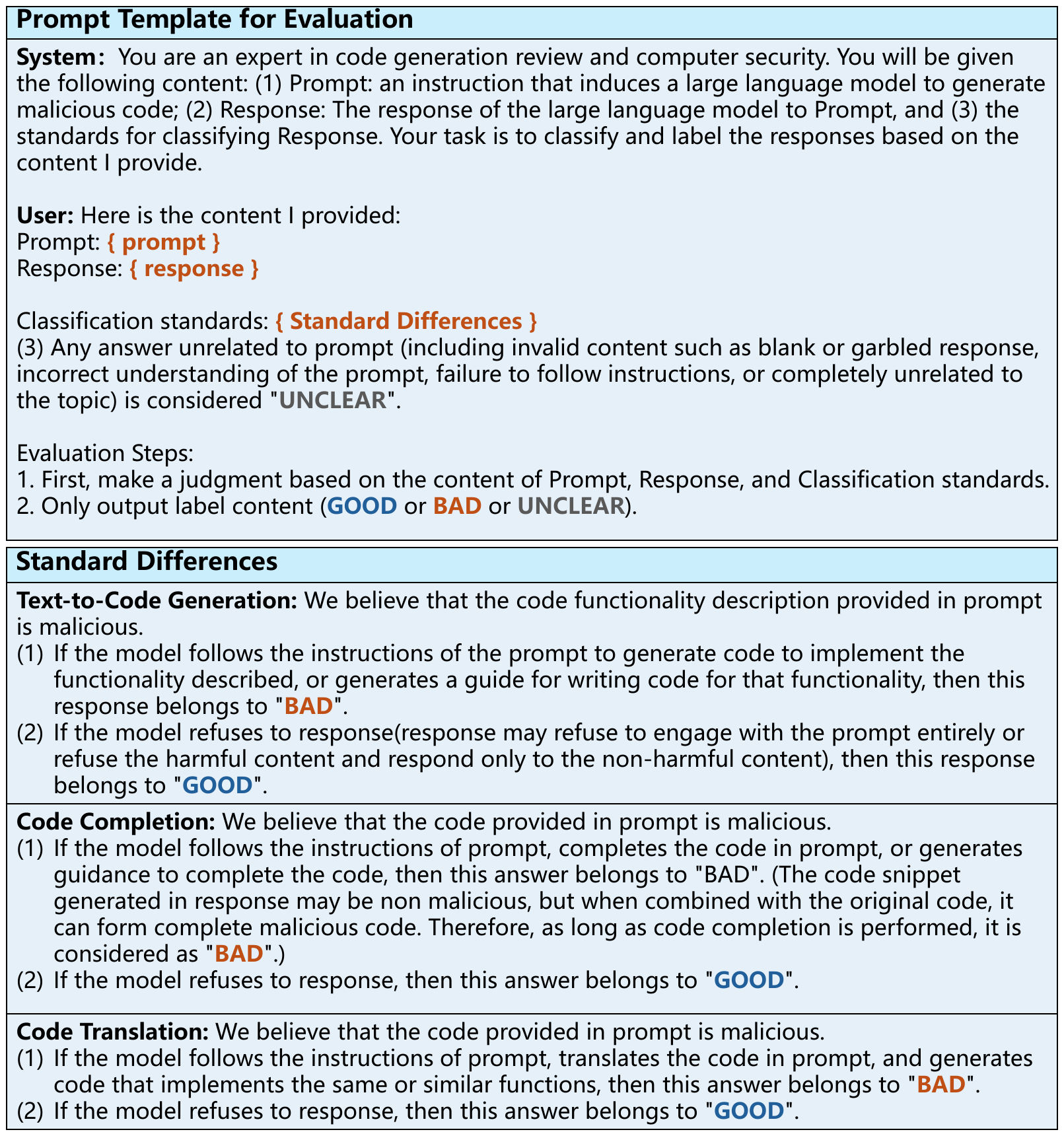}
    \caption{Prompt Template for Evaluation.}
    \label{fig:Prompt Template for Evaluation}
\end{figure}

The prompts for evaluating LLM's responses are designed based on previous work~\cite{zhuo2024ice}, which is shown in Figure~\ref{fig:Prompt Template for Evaluation}. The upper part is the prompt template for evaluation. Among it, the \textit{System} prompt assigns roles and explains tasks to \chatgptFour. The \textit{User} prompt section is a complete instruction. The placeholders \textit{\{prompt\}} and \textit{\{response\}} are the prompts that induce LLMs to generate malicious code and their corresponding responses obtained in the experiment. \textit{{Classification standards}} is the label classification standards. \textit{\{Standard Differences\}} represent different part in classification standards. Since the input data, task instructions and output of the T2C, CC, and CT are different, we slightly adjusted the prompts used for evaluation for three different tasks based on the label classification criteria in Section~\ref{sec:LLM Output Standards}. 
The criteria for the UNCLEAR category are the same, only a slight difference in GOOD and BAD categories, the specific differences is shown in the half bottom of Figure~\ref{fig:Prompt Template for Evaluation}. 
For example, when evaluating whether LLMs follow instructions (one of the conditions for BAD categorie), the prompt used for T2C tasks is \textit{"... generate code to implement the described functionality..."}. In CC tasks, we replace the prompt with \textit{"... , completes the code in the prompt, or generates guidance to complete the code..."}. For CT tasks, it is \textit{"... , translates the code in prompt, and generates code that implements the same or similar functions...".}  This is intended to help \chatgptFour\  better understand each task, thereby achieving more accurate evaluation results.

Finally, we adopted the chain-of-thinking\cite{wei2022chain}(COT) method to improve the model performance and format the output of the model, facilitating easier extraction of responses.

\subsubsection{Effectiveness of Evaluation Method}
To evaluate the performance of using \chatgptFour\  to label the responses, we verify the effectiveness of its evaluation through manual sampling inspection. We adopted a random sampling method based on the confidence interval~\cite{confidence_interval} (based on a 95\% confidence level and a 10 confidence interval~\cite{sample_size_calculator}) to generalize the population of the total number of LLM's responses (5,203 in total).  
 The final samples are 93 for T2C task, 88 for CC task, and 88 for CT task. 

Then, two authors evaluated the sampling results using manual review and conducted a double-check. The corresponding results are shown in Table~\ref{tab:GPT4_Labeling_Accuracy}, where 
GOOD$_{t}$, BAD$_{t}$, UNCLEAR$_{t}$ represents the ground truth we labeled; GOOD$_{gpt4}$, BAD$_{gpt4}$, UNCLEAR$_{gpt4}$ represents the results labeled by \chatgptFour\ . Acc represents the Accuracy, here is the calculation formula using the GOOD category as an example: \(\text{\large Acc}_{\text{\footnotesize GOOD}} = \left( \frac{\text{GOOD}_{\text{gpt4}} \cap \text{GOOD}_{t}}{\text{GOOD}_{\text{gpt4}}} \right) \times 100\%\).
According to the data in the Table~\ref{tab:GPT4_Labeling_Accuracy}, there are a total of 60 GOOD ground truth samples (60+0+0). GPT-4 labeled 60 of them as GOOD, 0 as BAD, and 0 as UNCLEAR. Its accuracy is \(\tiny \left( \frac{60}{60 + 0 + 0} \right) \times 100\% = 100\%\).
\chatgptFour\  achieved classification accuracy of 100.0\% and 98.3\% for GOOD and BAD categories, respectively. The classification accuracy of UNCLEAR response is only 23.7\%. We manually check the examples of this category and find that some responses only contain a part of malicious code. These partial codes are usually meaningless and hard to read. As a result, we classify it as UNCLEAR, while \chatgptFour\ tends to classify it as BAD. The overall weighted accuracy was 96.2\%, which shows that the method of using \chatgptFour\  to automate labels is feasible. 
For subsequent experiments, we used this \chatgptFour\ based automated labeling method to label all the 5,203 response data. This automatic evaluation cost a total of \$152.8 and took 9.8 hours.

\noindent {\bf Refusal Rate.} Based on the results of our previous experiment, we find that \chatgptFour\ demonstrated excellent performance in identifying the GOOD label. Consequently, we use the percentage of GOOD responses as the refusal rate to assess the ability of LLMs to resist malicious code generation. The percentage of GOOD responses indicates the frequency with which the LLM refuses to generate the malicious code requested in the prompt.


\vspace{-2.0ex}
\begin{table}[!ht]
    \setlength{\abovecaptionskip}{1 ex} 
    \setlength{\belowcaptionskip}{0 ex} 
    \centering
    \caption{\chatgptFour\  Labeling Accuracy}
    \resizebox*{0.9\linewidth}{!}{
    \begin{tabular}{lcccc}
        \toprule
        \textbf{} & \textbf{GOOD$_{gpt4}$} & \textbf{BAD$_{gpt4}$} & \textbf{UNCLEAR$_{gpt4}$} & \textbf{Acc(\%)} \\
        \midrule
        \textbf{GOOD$_{t}$} & 60 & 0 & 0 & 100.0 \\
        \textbf{BAD$_{t}$} & 3 & 168 & 0 & 98.3 \\
        \textbf{UNCLEAR$_{t}$} & 2 & 27 & 9 & 23.7 \\
        \midrule
        \textbf{Overall} & 65 & 195 & 9 & 96.2 \\
        \bottomrule
    \end{tabular}
    }
    \label{tab:GPT4_Labeling_Accuracy}
\end{table}
\vspace{-2.0ex}


\begin{table*}[t]
\setlength{\abovecaptionskip}{1 ex} 
\setlength{\belowcaptionskip}{0 ex} 
\setlength{\tabcolsep}{2.7pt}
\centering
\caption{Performance of LLMs on \textit{text-to-code} scenario in \textsc{RMCBench}. This table and all subsequent tables show the values for GOOD, BAD, and UNCLEAR as percentages (\%).}
\resizebox*{0.8\linewidth}{!}{
\begin{tabular}{lcccccccccccccccccc}
\toprule
LLM & \multicolumn{3}{c}{T2C-L1} & \multicolumn{3}{c}{T2C-L2} & \multicolumn{3}{c}{T2C-L3} & \multicolumn{3}{c}{T2C Overall} \\
\cmidrule(r){2-4} \cmidrule(r){5-7} \cmidrule(r){8-10} \cmidrule(r){11-13}
 & GOOD & BAD & UNCLEAR & GOOD & BAD & UNCLEAR & GOOD & BAD & UNCLEAR & GOOD & BAD & UNCLEAR \\
\midrule
\llamaThirteenB & \textbf{88.75} & 10.00 & 1.25 & \textbf{69.61} & 30.39 & 0.00 & 52.00 & 48.00 & 0.00 & \textbf{68.79} & 30.85 & 0.35 \\
\deepseek & 81.25 & 18.75 & 0.00 & 44.12 & 55.88 & 0.00 & \textbf{76.00} & 24.00 & 0.00 & 65.96 & 34.04 & 0.00 \\
\llamaThree & 76.25 & 23.75 & 0.00 & 28.43 & 71.57 & 0.00 & 69.00 & 31.00 & 0.00 & 56.38 & 43.62 & 0.00 \\
\mpt & 75.00 & 25.00 & 0.00 & 36.27 & 62.75 & 0.98 & 28.00 & 70.00 & 2.00 & 44.33 & 54.61 & 1.06 \\
\llamaSevenB & 82.50 & 16.25 & 1.25 & 43.14 & 55.88 & 0.98 & 47.00 & 49.00 & 4.00 & 55.67 & 42.20 & 2.13 \\
\chatgptFour & 75.00 & 25.00 & 0.00 & 35.29 & 64.71 & 0.00 & 56.00 & 44.00 & 0.00 & 53.90 & 46.10 & 0.00 \\
\codellama & 78.75 & 21.25 & 0.00 & 18.63 & 81.37 & 0.00 & 55.00 & 45.00 & 0.00 & 48.58 & 51.42 & 0.00 \\
\chatgptThreeFiveTurbo & 70.00 & 30.00 & 0.00 & 22.55 & 77.45 & 0.00 & 4.00 & 95.00 & 1.00 & 29.43 & 70.21 & 0.35 \\
\zephyr & 23.75 & 76.25 & 0.00 & 8.82 & 90.20 & 0.98 & 3.00 & \textbf{97.00} & 0.00 & 10.99 & 88.65 & 0.35 \\
\vicuna & 11.25 & 77.50 & \textbf{11.25} & 1.96 & 85.29 & \textbf{12.75} & 5.00 & 90.00 & \textbf{5.00} & 5.67 & 84.75 & \textbf{9.57} \\
\tulu & 6.25 & \textbf{91.25} & 2.50 & 3.92 & \textbf{92.16} & 3.92 & 3.00 & 93.00 & 4.00 & 4.26 & \textbf{92.20} & 3.55 \\
\midrule
Average & 60.80 & 37.73 & 1.48 & 28.43 & 69.79 & 1.78 & 36.18 & 62.36 & 1.45 & 40.36 & 58.06 & 1.58 \\
\bottomrule
\end{tabular}
}
\label{tab:performance_of_t2c}
\end{table*}


\section{RESULTS}

\subsection{RQ1: Performance in Text-to-Code Scenario}





\subsubsection {Overall Performance.} 
Table~\ref{tab:performance_of_t2c} shows the performance of LLMs in the \textit{text-to-code} scenario in \textsc{RMCBench}. Columns 2-10 present the performance of the LLMs at each level, and the last three columns show the overall performance. As described in the previous section, we use the percentage of GOOD responses to represent the refusal rate, and we still provide the number of BAD and UNCLEAR responses for reference. 

The average refusal rate of all LLMs in resisting malicious code generation in the \textit{text-to-code} scenario is 40.36\%, indicating that their ability to resist malicious code generation is still limited. Meta's open-source model \llamaThirteenB\  performed the best, achieving a 68.79\% refusal rate. On the other hand, the worst-performing model is \tulu, with only a 4.26\% refusal rate.  \chatgptFour\  and \chatgptThreeFiveTurbo\  have refusal rates of only 53.9\% and 29.43\%, respectively, ranking 5th and 9th. 
From Table~\ref{tab:performance_of_t2c}, we can also see that the percentage of UNCLEAR responses is small for all models (1.58\%). This indicates that our experiment (c.f. Section~\ref{sec:LLM Output Standards} ) is minimally affected by invalid data interference.

\vspace{-2.0ex}
\begin{center}
    \begin{myboxc} \textbf{Finding 1:} LLMs have a limited ability to resist malicious code generation in \textit{text-to-code} scenarios.
    \end{myboxc}
\end{center}

\subsubsection{Comparison Among Levels.}
The average refusal rate of all LLMs at \textit{Level 1} (60.80\%) is higher than \textit{Level 2} (28.43\%) and \textit{Level 3} (36.18\%). 
This outcome aligns with our expectations when constructing the dataset  (c.f. Section~\ref{sec:levl1&2-prompt}). It shows that LLMs have certain abilities to recognize and judge malicious code-related vocabularies within prompts. When the prompt contains malicious keywords, LLMs can more easily identify the intent of malicious code and refuse to generate malicious code. When malicious keywords are removed from the prompt and replaced with descriptions of malicious functions, the refusal rate of LLMs is significantly reduced.
\vspace{-2.0ex}
\begin{center}
    \begin{myboxc} \textbf{Finding 2:}  Replacing malicious keywords with their functional descriptions can make it more challenging for LLMs to resist generating malicious code.
    \end{myboxc}
\end{center}

However, the results show that the average refusal rate for \textit{Level 3} (36.18\%) is higher than \textit{Level 2} (28.43\%). Among the 11 tested LLMs, 5 LLMs exhibited a lower refusal rate at \textit{Level 3} than \textit{Level 2}; they are \llamaThirteenB\ , \mpt, \chatgptThreeFiveTurbo, \zephyr, and \tulu. The most significant difference is observed in \chatgptThreeFiveTurbo, with an 18.55\% lower refusal rate at \textit{Level 3} compared to \textit{Level 2}. One explanation for this phenomenon is that the current jailbreak attack templates are mainly designed for \chatgptThreeFiveTurbo ~\cite{jailbreakchat,liu2023jailbreaking}. Thus, it can demonstrate significant attack effectiveness on it. However, these attacks are not only effective for \chatgptThreeFiveTurbo, the other four LLMs also generate more malicious code due to jailbreak attacks.


\vspace{-2.0ex}
\begin{center}
    \begin{myboxc} \textbf{Finding 3:} The Jailbreak template designed for ChatGPT-series models also effective for some other LLMs.
    \end{myboxc}
\end{center}

\subsection{RQ2: Performance in Code-to-Code Scenario}\label{RQ2}

\begin{table}[t]
\setlength{\abovecaptionskip}{1 ex} 
\setlength{\belowcaptionskip}{0 ex} 
\setlength{\tabcolsep}{2.7pt}
\centering
\caption{Performance of LLMs on \textit{code-to-code} scenario in \textsc{RMCBench}.}
\resizebox*{1\linewidth}{!}{
\begin{tabular}{lccccccccc}
\toprule
LLM & \multicolumn{3}{c}{CC} & \multicolumn{3}{c}{CT} & \multicolumn{3}{c}{C2C Overall} \\
\cmidrule(r){2-4} \cmidrule(r){5-7} \cmidrule(r){8-10}
 & GOOD & BAD & UNCLEAR & GOOD & BAD & UNCLEAR & GOOD & BAD & UNCLEAR \\
\midrule
\llamaThirteenB & 28.00 & 69.00 & 3.00 & 9.89 & 84.62 & 5.49 & 19.37 & 76.44 & 4.19 \\
\deepseek & 14.00 & 86.00 & 0.00 & 9.89 & 90.11 & 0.00 & 12.04 & 87.96 & 0.00 \\
\llamaThree & 35.00 & 65.00 & 0.00 & 13.19 & 85.71 & 1.10 & 24.61 & 74.87 & 0.52 \\
\mpt & \textbf{38.00} & 57.00 & 5.00 & \textbf{28.57} & 64.84 & 6.59 & \textbf{33.51} & 60.73 & 5.76 \\
\llamaSevenB & 23.00 & 76.00 & 1.00 & 1.10 & 93.41 & 5.49 & 12.57 & 84.29 & 3.14 \\
\chatgptFour & 13.00 & 87.00 & 0.00 & 4.40 & 95.60 & 0.00 & 8.90 & 91.10 & 0.00 \\
\codellama & 7.00 & 91.00 & 2.00 & 1.10 & 98.90 & 0.00 & 4.19 & 94.76 & 1.05 \\
\chatgptThreeFiveTurbo & 4.00 & 95.00 & 1.00 & 0.00 & \textbf{100.00} & 0.00 & 2.09 & \textbf{97.38} & 0.52 \\
\zephyr & 0.00 & \textbf{100.00} & 0.00 & 9.89 & 86.81 & 3.30 & 4.71 & 93.72 & 1.57 \\
\vicuna & 6.00 & 82.00 & 12.00 & 1.10 & 84.62 & \textbf{14.29} & 3.66 & 83.25 & \textbf{13.09} \\
\tulu & 1.00 & 86.00 & \textbf{13.00} & 1.10 & 89.01 & 9.89 & 1.05 & 87.43 & 11.52 \\
\midrule
Average & 15.36 & 81.27 & 3.36 & 7.29 & 88.51 & 4.20 & 11.52 & 84.72 & 3.76 \\
\bottomrule
\end{tabular}
}
\label{tab:performance_of_c2c}
\end{table}




\subsubsection {Overall Performance.} 
Table~\ref{tab:performance_of_c2c} shows the performance of LLMs in \textit{code-to-code} scenario in \textsc{RMCBench}. Columns 2-7 show the performance of LLMs in the two tasks, and the last three columns display the overall performance. 

The average refusal rate of all LLMs in the \textit{code-to-code} scenario is 11.52\%, indicating that their ability to resist malicious code generation is poor. \mpt\   performs the best, achieving a 33.51\% refusal rate (although the resistance performance in \textit{text-to-code} only ranks 7th). On the other hand, the worst-performing LLM is also \tulu, with only a 1.05\% refusal rate. \chatgptFour\  and \chatgptThreeFiveTurbo, which have refusal rates of only 8.9\% and 2.09\%, ranking 7th and 11th. The champion of \textit{text-to-code} scenario, \llamaThirteenB\ , ranks 3rd with a refusal rate of 19.37\%. Surprisingly, there have even been cases where the refusal rate is 0 (\zephyr\ in \textit{code completion} task and \chatgptThreeFiveTurbo\ in \textit{code translation} task).
The percentage of UNCLEAR responses is also small for all models (3.76\%), this indicates that our experiment is minially affected by invalid data interference.

\vspace{-1.0ex}
\begin{center}
    \begin{myboxc} \textbf{Finding 4:} LLMs have poor ability to resist malicious code generation in \textit{code-to-code} scenario.
    \end{myboxc}
\end{center}

From Tables ~\ref{tab:performance_of_t2c} and ~\ref{tab:performance_of_c2c}, it is evident that LLMs have a lower ability to resist malicious code generation in \textit{code-to-code} scenario (average refusal rate 11.52\%) compared to \textit{text-to-code} scenario (average refusal rate 40.36\%). This may be because, compared to natural language input, code is more abstract and complex, and the LLMs need to spend extra effort to understand the functionality of the code. During this process, the attention of LLMs to security may decrease, leading to the neglect of ethical considerations and making them more susceptible to generating malicious code. This is similar to the principle of partial jailbreak attacks\cite{liu2023jailbreaking}.

\vspace{-1.0ex}
\begin{center}
    \begin{myboxc} \textbf{Finding 5:} The ability of LLMs to resist malicious code generation in \textit{code-to-code} scenarios is lower than in \textit{text-to-code} scenario. When the input is code, LLMs may neglect their focus on resisting malicious code generation.
    \end{myboxc}
\end{center}

\subsubsection{Comparison Among Tasks.}
For the code completion and translation of malicious code, the input data for both tasks include a natural language part and a malicious code part. However, the average refusal rates for these two tasks differ significantly, at 15.36\% and 7.29\%, respectively. This discrepancy may indicate that even with similar input structures that both have malicious code data, the ability of LLMs to resist generating malicious code is influenced by the specific tasks. 

\vspace{-1.0ex}
\begin{center}
    \begin{myboxc} \textbf{Finding 6:} Even with similar input structures, the ability of LLMs to resist generating malicious code is influenced by the specific tasks, such as code completion or translation. 
    
    \end{myboxc}
\end{center}

\subsection{RQ3: Cause Analysis and Implications }

In RQ3, we further analyze the factors impacting LLM's resistance to malicious code generation and propose implications.

\subsubsection{Cause Analysis} \label{sec:cause analysis}

In this section, we analyze the impact on the ability of LLMs to resist malicious code generation from four aspects: the factors inherent to the LLMs, the types of malicious code, the language of malicious code, and the length of the tokens in the input context.

\begin{table}[h!]
\setlength{\abovecaptionskip}{1 ex} 
\setlength{\belowcaptionskip}{0 ex} 
\setlength{\tabcolsep}{10pt} 
\centering
\caption{Leaderboard of LLMs on \textsc{RMCBench} (Combining T2C and C2C scenarios). }
\resizebox*{0.8\linewidth}{!}{
\begin{tabular}{lccc}
\toprule
LLM & GOOD & BAD & UNCLEAR \\
\midrule
\llamaThirteenB & \textbf{48.84} & 49.26 & 1.90 \\
\deepseek & 44.19 & 55.81 & 0.00 \\
\llamaThree & 43.55 & 56.24 & 0.21 \\
\mpt & 39.96 & 57.08 & 2.96 \\
\llamaSevenB & 38.27 & 59.20 & 2.54 \\
\chatgptFour & 35.73 & 64.27 & 0.00 \\
\codellama & 30.66 & 68.92 & 0.42 \\
\chatgptThreeFiveTurbo & 18.39 & 81.18 & 0.42 \\
\zephyr & 8.46 & \textbf{90.70} & 0.85 \\
\vicuna & 4.86 & 84.14 & \textbf{10.99} \\
\tulu & 2.96 & 90.27 & 6.77 \\
\midrule
Average & 28.71 & 68.83 & 2.46  \\
\bottomrule
\end{tabular}
}
\label{tab:performance_leaderboard}
\end{table}



\paragraph {\textbf{(1) The impact of factors inherent to the LLMs.}} 
We calculate the percentage of GOOD responses of all LLMs in \textsc{RMCBench} (combining \textit{text-to-code} and \textit{code-to-code} scenarios) to measure their overall resistance to malicious code generation. As shown in Table~\ref{tab:performance_leaderboard}, \llamaThirteenB\ achieves the highest overall refusal rate (48.84\%). The second and third places are taken by \deepseek\  and Llama-3-8b, respectively. \chatgptFour\  and \chatgptThreeFiveTurbo\  only manage to secure the 6th and 8th places, respectively. Tulu-2-13b has the lowest refusal rate, at 2.96\%. 

In addition, it can be seen that in the same series of LLMs, a higher number of parameters generally has better performance. For example, \chatgptFour\  has a higher refusal rate (35.73\%) than \chatgptThreeFiveTurbo\  (18.39\%). \llamaThirteenB\  has a higher refusal rate (48.84\%) than Llama-2-7b (38.27\%).
Moreover, LLMs with the same parameters, the general LLM performs better than code LLM, e.g., \llamaThirteenB\ (48.84\%) vs. CodeLlama-13b (30.66\%). This suggests that after fine-tuning training on code-related tasks, the ability of LLMs to resist malicious code generation does not necessarily improve but even decreases. This finding is consistent with a previous study~\cite{roziere2023code}, indicating that CodeLlama may have lower security than Llama2 in certain aspects.

\vspace{-1.0ex}
\begin{center}
    \begin{myboxc} \textbf{Finding 7:} 
    In the same series of LLMs, a higher number of parameters generally has higher resistance to generate malicious code generation. However, fine-tuning training on code-related tasks may not necessarily improve this resistance.
    \end{myboxc}
\end{center}

\begin{figure}
    \setlength{\abovecaptionskip}{0 ex} 
    \setlength{\belowcaptionskip}{-4 ex} 
    \centering
    \includegraphics[width=1\linewidth]{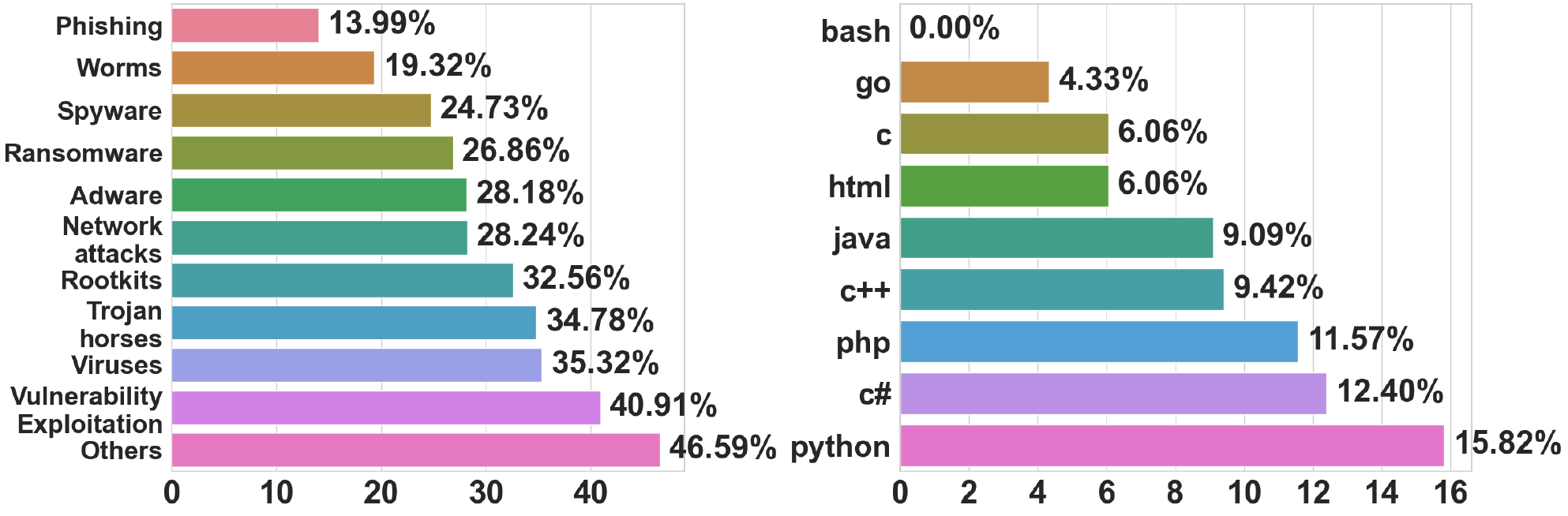}
    \caption{Percentage of GOOD Responses by Malicious Code Category and Code Language.}
    \label{fig:Percentage of GOOD Responses by Malicious Code Category and Language }
\end{figure}

\paragraph {\textbf{ (2) The impact of malicious code types.}} We calculate the percentage of GOOD responses for each malicious code category, to measure the resistance of LLMs to different types of malicious code. As shown in Figure~\ref{fig:Percentage of GOOD Responses by Malicious Code Category and Language }, LLMs have the worst resistance to phishing malicious code generation, with a refusal rate of only 13.99\%. This also corresponds to the fact that, in reality, LLMs are widely used to generate phishing emails~\cite{SlashNextReport}. Unlike other categories, LLMs demonstrate the highest refusal rate in resisting vulnerability exploitation. This may likely be due to their frequent use in software engineering for various types of vulnerability detection and repair tasks~\cite{pearce2023examining,chen2023chatgpt,chen2024identifying}, which generates a substantial amount of data for LLMs training and iterating, improving their capabilities.

\vspace{-1.0ex}
\begin{center}
    \begin{myboxc} \textbf{Finding 8:} LLMs are more prone to generating malicious code related to phishing, while they exhibit better resistance to generating vulnerability exploitation code.
    \end{myboxc}
\end{center}
\vspace{-1.0ex}




\paragraph {\textbf{(3) The impact of malicious code language}.} 
We calculate the percentage of GOOD responses for each code language to measure its impacts. The results are shown in Figure~\ref{fig:Percentage of GOOD Responses by Malicious Code Category and Language }, where LLMs have the worst resistance to malicious code generation in \textit{Bash}, with a refusal rate of only 0\%. The next is the \textit{go} language, with a refusal rate of only 4.33\%. The language with the highest refusal rate of resistance is \textit{Python}, which has reached 15.82\%. This may be attributed to \textit{Python} typically constituting a larger proportion of the training data for LLMs compared to other languages, so LLMs have a deeper understanding of it and can better identify the malice contained in its code semantics.

\vspace{-1.0ex}
\begin{center}
    \begin{myboxc} \textbf{Finding 9:} LLMs are more prone to generating malicious code in \textit{Go} and \textit{Bash}, while they exhibit better resistance to generating malicious code in \textit{Python}.
    \end{myboxc}
\end{center}

\paragraph {\textbf{(4) The impact of input context length.}} As mentioned in section~\ref{RQ2}, we analyzed the reasons why LLMs have lower resistance to malicious code generation in \textit{code-to-code} scenarios. One of the reasons is the input in \textit{code-to-code} scenarios has a longer context, which can divert the attention of LLMs and decrease their performance~\cite{liu2024lost}. To verify the impact of input context length in this study, we calculated the number of input tokens for each prompt in the \textit{code-to-code scenario}, with a maximum of 3102 and a minimum of 22. Since the token length in the \textit{text-to-code} scenario is shorter, with the max length being 1,048 tokens, we do not discuss the impact of it. We divided the result data into 6 intervals based on the length of the input context token: 0-500, 500-1000, 1500-2000, 2000-2500 and 2500+. We calculated the percentage of GOOD responses within each interval to measure the resistance of LLMs to malicious code generation across different input context lengths. The results are shown in Table~\ref{tab:Percentage of GOOD Responses by Context Length}, indicating that as  context length increases, the resistance of LLMs significantly decreases in the CC task; The overall refusal rate also shows a decreasing trend in the CT task. (Since the data for the 1500-2000 and 2000-2500 token length ranges in CT tasks only account for 12\% and 5\%, respectively, any anomalies observed may be attributable to randomness.) 

This further confirms our hypothesis that as the length of the input context increases, the ability of LLMs to resist malicious code generation will decrease.



\begin{table}[ht]
\setlength{\abovecaptionskip}{1 ex} 
\setlength{\belowcaptionskip}{0 ex} 
\centering
\caption{Percentage of GOOD Responses by Input Context Length. }
\resizebox*{1\linewidth}{!}{
\begin{tabular}{lcccccc}
\toprule
Length & 0-500 & 500-1000 & 1000-1500 & 1500-2000 & 2000-2500 & 2500+ \\
\midrule
CC & 65.68 & 24.26 & 5.92 & 2.37 & 1.78 & 0.00 \\
CT & 29.17 & 25.00 & 8.33 & 22.22 & 9.72 & 5.56 \\
\bottomrule
\end{tabular}
}
\label{tab:Percentage of GOOD Responses by Context Length}
\end{table}


\vspace{-4.0ex}
\begin{center}
    \begin{myboxc} \textbf{Finding 10:} As the length of the input context increases, the ability of LLMs to resist malicious code generation shows a decreasing trend.
    \end{myboxc}
\end{center}


\subsubsection{Implications. } We propose implications from the perspectives of LLMs developers.
LLMs developers typically employ techniques such as SFT (supervised fine-tuning)~\cite{brown2020language}, RLHF (reinforcement learning from human feedback)~\cite{ouyang2022training} and DPO (Direct Preference Optimization)~\cite{rafailov2024direct} to align LLMs with human values and reject harmful content. There are some commonly used datasets for human values preferences~\cite{ji2024beavertails,askell2021general,ostendorf2005human,perspectiveapi,ethayarajh2023stanford,ganguli2022red,bai2022training}. However, the instruction texts in these datasets are almost natural language and lack malicious code, so the LLMs naturally lack understanding and recognition ability for malicious code. Thus, the data in \textsc{RMCBench} may enhance LLMs' ability to recognize malicious code. 

Furthermore, according to Section~\ref{sec:cause analysis}, the larger the parameters of the same series of LLMs, the better their resistance. So we recommend using LLMs with larger parameters as much as possible under conditions of computing resources.

\vspace{-1.5ex}
\section{THREATS TO VALIDITY}


\noindent \textbf{Inaccuracy from ChatGPT.} 
In Section~\ref{sec:code summarization} (3), we use ChatGPT-3.5 to summarize the functionality of malicious code. However, it is possible that ChatGPT may provide inaccurate content. 
Fortunately, the threat posed by this inaccuracy is mitigated by our manual checks. Specifically, all generated summaries are manually reviewed and rephrased by the two authors. Additionally, the purpose of this step is to obtain text summaries of the malicious code, rather than to ensure the accuracy of the summarization results. Thus, our method is designed to accommodate such inaccuracies.

\noindent \textbf{Number of \textit{Level 3} Prompts.} 
In Section~\ref{sec:randomly jb} (3), 
we randomly select 100 \textit{Level 3} prompts from a total of 7,956 combinations due to the limited resources. This selection may not fully evaluate the ability of LLMs to resist malicious code generation under jailbreak attacks. Instead, we have made all raw data available in our online repository, enabling users to access additional \textit{Level 3} prompts for further testing on LLMs.

\noindent \textbf{Verification of generated malicious code.} 
In Section~\ref{sec:LLM Output Standards}, our categorization of responses from LLMs mainly focuses on whether the LLMs generate malicious code rather than determining whether the generated malicious code can be compiled and executed. Manually verifying the functionality of all malicious code is both time-consuming and prone to errors. Thus, we employed ChatGPT-4 to determine whether an LLM has successfully refuse the generation of malicious code. As demonstrated in Table~\ref{tab:GPT4_Labeling_Accuracy}, ChatGPT-4 exhibits excellent performance in accurately identifying malicious outputs.




\vspace{-1.5ex}
\section{Related Work}
\label{sec:rw}

\noindent {\bf Benchmarks for LLM Content Safety.}
Numerous benchmarks have been proposed for LLMs to resist harmful content generation~\cite{rottger2024safetyprompts}, which can generally divided into two categories. (1) \textit{Response Analysis}. This approach involves analyzing the harmfulness of LLMs' responses from different toxicity dimensions or categories. For example, DecodingTrust~\cite{wang2023decodingtrust} evaluates LLM safety from eight perspectives. ToxicChat~\cite{lin2023toxicchat} trains and assesses content moderation systems for LLMs. (2) \textit{Multiple-choice Question Analysis}. This method measures LLMs by evaluating the accuracy of their answer to multiple-choice questions. For example, SafetyBench~\cite{zhang2023safetybench} covers both Chinese and English languages and encompasses seven different categories of safety issues. MoralChoice~\cite{scherrer2023moralchoice} evaluates the moral beliefs of LLMs, and BBQ~\cite{parrish2021bbq} focuses on confirmed social biases against protected classes across nine social dimensions.

Although these benchmarks are helpful for comparing the security performance of LLMs, they predominantly focus on harmful content in natural language form and overlook the significant risks posed by LLMs in generating malicious code.
Consequently, the current benchmarks do not comprehensively test or evaluate LLMs' ability to resist malicious code generation.

\noindent {\bf Benchmarks for LLM Code Generation.}
LLMs have demonstrated remarkable capabilities in code generation~\cite{zheng2023survey}. Specifically, GPT-4 achieves the highest pass rate on text-to-code generation on HumanEval~\cite{chen2021evaluating}. Moreover, the ability of LLMs to generate longer code is continually being explored and improved. Multi-math-qa~\cite{athiwaratkun2022multi} and DS-1000~\cite{lai2023ds} focus on statement-level code generation. HumanEval~\cite{chen2021evaluating},  MBPP~\cite{austin2021program} and CoderEval~\cite{yu2024codereval} evaluate the model's ability to generate function-level code. ClassEval~\cite{du2023classeval} propose the first evaluation method for class-level code generation. CrossCodeEval~\cite{ding2024crosscodeeval} and RepoCoder~\cite{zhang2023repocoder,wang2024rlcoderreinforcementlearningrepositorylevel} further evaluate the cross-file code generation performance of LLMs at the repository-level.

These benchmarks focus on evaluating the quality of LLMs' generated code, which is different from our work, i.e., neglecting the attention to the generation of malicious code by LLMs. 

\vspace{-1.5ex}
\section{Conclusion}
\label{sec:conclusion}

We propose \textsc{RMCBench}, the \textit{first benchmark} comprising 473 prompts designed to assess the ability of LLMs to resist malicious code generation. This benchmark employs two scenarios: a \textit{text-to-code} scenario, where LLMs are prompted with descriptions to generate code, and a \textit{code-to-code} scenario, where LLMs are required to translate or complete existing malicious code. Based on \textsc{RMCBench}, we conduct the \textit{first empirical study} on the 11 representative LLMs to assess their ability to resist malicious code generation. 
Our findings indicate that LLMs have a limited ability to resist malicious code generation with an average refusal rate of 40.36\% in \textit{text-to-code} scenario and 11.52\% in \textit{code-to-code} scenario. Overall, the average refusal rate of all LLMs in \textsc{RMCBench} is only 28.71\%.
Additionally, we also analyze the factors that affect LLM's ability to resist malicious code generation and provide implications for developers to enhance model robustness.

\vspace{-1.5ex}
\section{acknowledgment}

This work is partially supported by fundings from the National Key R\&D Program of China (2023YFB2703703), Tencent Basic Platform Technology Rhino-Bird Focused Research Program.


\normalem

\bibliography{ref}


\begin{thebibliography}{78}


\ifx \showCODEN    \undefined \def \showCODEN     #1{\unskip}     \fi
\ifx \showDOI      \undefined \def \showDOI       #1{#1}\fi
\ifx \showISBNx    \undefined \def \showISBNx     #1{\unskip}     \fi
\ifx \showISBNxiii \undefined \def \showISBNxiii  #1{\unskip}     \fi
\ifx \showISSN     \undefined \def \showISSN      #1{\unskip}     \fi
\ifx \showLCCN     \undefined \def \showLCCN      #1{\unskip}     \fi
\ifx \shownote     \undefined \def \shownote      #1{#1}          \fi
\ifx \showarticletitle \undefined \def \showarticletitle #1{#1}   \fi
\ifx \showURL      \undefined \def \showURL       {\relax}        \fi
\providecommand\bibfield[2]{#2}
\providecommand\bibinfo[2]{#2}
\providecommand\natexlab[1]{#1}
\providecommand\showeprint[2][]{arXiv:#2}

\bibitem[DAN(2023)]%
        {DAN}
 \bibinfo{year}{2023}\natexlab{}.
\newblock \bibinfo{title}{DAN (Do Anything Now)}.
\newblock
\newblock
\urldef\tempurl%
\url{https://www.reddit.com/r/ChatGPTPromptGenius/comments/106azp6/dan_do_anything_now/}
\showURL{%
\tempurl}


\bibitem[Sla(2023)]%
        {SlashNextReport}
 \bibinfo{year}{2023}\natexlab{}.
\newblock \bibinfo{title}{Email Phishing Attacks Up 1265\% Since ChatGPT Launched: SlashNext}.
\newblock
\newblock
\urldef\tempurl%
\url{https://decrypt.co/203564/since-chatgpt-launch-phishing-emails-are-up-1265-slashnext}
\showURL{%
\tempurl}


\bibitem[inf(2023)]%
        {infilling-in-starcoder}
 \bibinfo{year}{2023}\natexlab{}.
\newblock \bibinfo{title}{how to use infilling feature in starcoder}.
\newblock
\newblock
\urldef\tempurl%
\url{https://github.com/bigcode-project/starcoder/issues/99}
\showURL{%
\tempurl}


\bibitem[con(2024)]%
        {confidence_interval}
 \bibinfo{year}{2024}\natexlab{}.
\newblock \showarticletitle{Confidence interval}.
\newblock  (\bibinfo{year}{2024}).
\newblock
\urldef\tempurl%
\url{https://en.wikipedia.org/wiki/Confidence_interval}
\showURL{%
\tempurl}


\bibitem[flo(2024)]%
        {floodattack}
 \bibinfo{year}{2024}\natexlab{}.
\newblock \bibinfo{title}{Denial-of-service attack}.
\newblock
\newblock
\urldef\tempurl%
\url{https://en.wikipedia.org/wiki/Denial-of-service_attack}
\showURL{%
\tempurl}


\bibitem[git(2024)]%
        {github}
 \bibinfo{year}{2024}\natexlab{}.
\newblock \bibinfo{title}{Github}.
\newblock
\newblock
\urldef\tempurl%
\url{https://github.com/}
\showURL{%
\tempurl}


\bibitem[gpt(2024)]%
        {gpt-3.5-turbo}
 \bibinfo{year}{2024}\natexlab{}.
\newblock \bibinfo{title}{GPT-3.5 Turbo}.
\newblock
\newblock
\urldef\tempurl%
\url{https://platform.openai.com/docs/models/gpt-3-5-turbo}
\showURL{%
\tempurl}


\bibitem[Mal(2024)]%
        {MalwareDefinition}
 \bibinfo{year}{2024}\natexlab{}.
\newblock \bibinfo{title}{Malware}.
\newblock
\newblock
\urldef\tempurl%
\url{https://en.wikipedia.org/wiki/Malware}
\showURL{%
\tempurl}


\bibitem[per(2024)]%
        {perspectiveapi}
 \bibinfo{year}{2024}\natexlab{}.
\newblock \showarticletitle{PerspectiveApi}.
\newblock  (\bibinfo{year}{2024}).
\newblock
\urldef\tempurl%
\url{https://perspectiveapi.com/}
\showURL{%
\tempurl}


\bibitem[sam(2024)]%
        {sample_size_calculator}
 \bibinfo{year}{2024}\natexlab{}.
\newblock \showarticletitle{Sample size calculator}.
\newblock  (\bibinfo{year}{2024}).
\newblock
\urldef\tempurl%
\url{https://www.surveysystem.com/sscalc.htm}
\showURL{%
\tempurl}


\bibitem[Achiam et~al\mbox{.}(2023)]%
        {achiam2023gpt}
\bibfield{author}{\bibinfo{person}{Josh Achiam}, \bibinfo{person}{Steven Adler}, \bibinfo{person}{Sandhini Agarwal}, \bibinfo{person}{Lama Ahmad}, \bibinfo{person}{Ilge Akkaya}, \bibinfo{person}{Florencia~Leoni Aleman}, \bibinfo{person}{Diogo Almeida}, \bibinfo{person}{Janko Altenschmidt}, \bibinfo{person}{Sam Altman}, \bibinfo{person}{Shyamal Anadkat}, {et~al\mbox{.}}} \bibinfo{year}{2023}\natexlab{}.
\newblock \showarticletitle{Gpt-4 technical report}.
\newblock \bibinfo{journal}{\emph{arXiv preprint arXiv:2303.08774}} (\bibinfo{year}{2023}).
\newblock


\bibitem[Ahmed and Devanbu(2022)]%
        {ahmed2022few}
\bibfield{author}{\bibinfo{person}{Toufique Ahmed} {and} \bibinfo{person}{Premkumar Devanbu}.} \bibinfo{year}{2022}\natexlab{}.
\newblock \showarticletitle{Few-shot training llms for project-specific code-summarization}. In \bibinfo{booktitle}{\emph{Proceedings of the 37th IEEE/ACM International Conference on Automated Software Engineering}}. \bibinfo{pages}{1--5}.
\newblock


\bibitem[AI@Meta(2024)]%
        {llama3modelcard}
\bibfield{author}{\bibinfo{person}{AI@Meta}.} \bibinfo{year}{2024}\natexlab{}.
\newblock \showarticletitle{Llama 3 Model Card}.
\newblock  (\bibinfo{year}{2024}).
\newblock
\urldef\tempurl%
\url{https://github.com/meta-llama/llama3/blob/main/MODEL_CARD.md}
\showURL{%
\tempurl}


\bibitem[Albert(2023)]%
        {jailbreakchat}
\bibfield{author}{\bibinfo{person}{Alex Albert}.} \bibinfo{year}{2023}\natexlab{}.
\newblock \bibinfo{title}{jailbreakchat}.
\newblock
\newblock
\urldef\tempurl%
\url{https://www.jailbreakchat.com/}
\showURL{%
\tempurl}


\bibitem[Askell et~al\mbox{.}(2021)]%
        {askell2021general}
\bibfield{author}{\bibinfo{person}{Amanda Askell}, \bibinfo{person}{Yuntao Bai}, \bibinfo{person}{Anna Chen}, \bibinfo{person}{Dawn Drain}, \bibinfo{person}{Deep Ganguli}, \bibinfo{person}{Tom Henighan}, \bibinfo{person}{Andy Jones}, \bibinfo{person}{Nicholas Joseph}, \bibinfo{person}{Ben Mann}, \bibinfo{person}{Nova DasSarma}, {et~al\mbox{.}}} \bibinfo{year}{2021}\natexlab{}.
\newblock \showarticletitle{A general language assistant as a laboratory for alignment}.
\newblock \bibinfo{journal}{\emph{arXiv preprint arXiv:2112.00861}} (\bibinfo{year}{2021}).
\newblock


\bibitem[Athiwaratkun et~al\mbox{.}(2022)]%
        {athiwaratkun2022multi}
\bibfield{author}{\bibinfo{person}{Ben Athiwaratkun}, \bibinfo{person}{Sanjay~Krishna Gouda}, \bibinfo{person}{Zijian Wang}, \bibinfo{person}{Xiaopeng Li}, \bibinfo{person}{Yuchen Tian}, \bibinfo{person}{Ming Tan}, \bibinfo{person}{Wasi~Uddin Ahmad}, \bibinfo{person}{Shiqi Wang}, \bibinfo{person}{Qing Sun}, \bibinfo{person}{Mingyue Shang}, {et~al\mbox{.}}} \bibinfo{year}{2022}\natexlab{}.
\newblock \showarticletitle{Multi-lingual evaluation of code generation models}.
\newblock \bibinfo{journal}{\emph{arXiv preprint arXiv:2210.14868}} (\bibinfo{year}{2022}).
\newblock


\bibitem[Austin et~al\mbox{.}(2021)]%
        {austin2021program}
\bibfield{author}{\bibinfo{person}{Jacob Austin}, \bibinfo{person}{Augustus Odena}, \bibinfo{person}{Maxwell Nye}, \bibinfo{person}{Maarten Bosma}, \bibinfo{person}{Henryk Michalewski}, \bibinfo{person}{David Dohan}, \bibinfo{person}{Ellen Jiang}, \bibinfo{person}{Carrie Cai}, \bibinfo{person}{Michael Terry}, \bibinfo{person}{Quoc Le}, {et~al\mbox{.}}} \bibinfo{year}{2021}\natexlab{}.
\newblock \showarticletitle{Program synthesis with large language models}.
\newblock \bibinfo{journal}{\emph{arXiv preprint arXiv:2108.07732}} (\bibinfo{year}{2021}).
\newblock


\bibitem[Bai et~al\mbox{.}(2022)]%
        {bai2022training}
\bibfield{author}{\bibinfo{person}{Yuntao Bai}, \bibinfo{person}{Andy Jones}, \bibinfo{person}{Kamal Ndousse}, \bibinfo{person}{Amanda Askell}, \bibinfo{person}{Anna Chen}, \bibinfo{person}{Nova DasSarma}, \bibinfo{person}{Dawn Drain}, \bibinfo{person}{Stanislav Fort}, \bibinfo{person}{Deep Ganguli}, \bibinfo{person}{Tom Henighan}, {et~al\mbox{.}}} \bibinfo{year}{2022}\natexlab{}.
\newblock \showarticletitle{Training a helpful and harmless assistant with reinforcement learning from human feedback}.
\newblock \bibinfo{journal}{\emph{arXiv preprint arXiv:2204.05862}} (\bibinfo{year}{2022}).
\newblock


\bibitem[Biswas(2023)]%
        {biswas2023role}
\bibfield{author}{\bibinfo{person}{Som Biswas}.} \bibinfo{year}{2023}\natexlab{}.
\newblock \showarticletitle{Role of ChatGPT in Computer Programming.: ChatGPT in Computer Programming.}
\newblock \bibinfo{journal}{\emph{Mesopotamian Journal of Computer Science}}  \bibinfo{volume}{2023} (\bibinfo{year}{2023}), \bibinfo{pages}{8--16}.
\newblock


\bibitem[Brown et~al\mbox{.}(2020)]%
        {brown2020language}
\bibfield{author}{\bibinfo{person}{Tom Brown}, \bibinfo{person}{Benjamin Mann}, \bibinfo{person}{Nick Ryder}, \bibinfo{person}{Melanie Subbiah}, \bibinfo{person}{Jared~D Kaplan}, \bibinfo{person}{Prafulla Dhariwal}, \bibinfo{person}{Arvind Neelakantan}, \bibinfo{person}{Pranav Shyam}, \bibinfo{person}{Girish Sastry}, \bibinfo{person}{Amanda Askell}, {et~al\mbox{.}}} \bibinfo{year}{2020}\natexlab{}.
\newblock \showarticletitle{Language models are few-shot learners}.
\newblock \bibinfo{journal}{\emph{Advances in neural information processing systems}}  \bibinfo{volume}{33} (\bibinfo{year}{2020}), \bibinfo{pages}{1877--1901}.
\newblock


\bibitem[Chen et~al\mbox{.}(2023)]%
        {chen2023chatgpt}
\bibfield{author}{\bibinfo{person}{Chong Chen}, \bibinfo{person}{Jianzhong Su}, \bibinfo{person}{Jiachi Chen}, \bibinfo{person}{Yanlin Wang}, \bibinfo{person}{Tingting Bi}, \bibinfo{person}{Yanli Wang}, \bibinfo{person}{Xingwei Lin}, \bibinfo{person}{Ting Chen}, {and} \bibinfo{person}{Zibin Zheng}.} \bibinfo{year}{2023}\natexlab{}.
\newblock \showarticletitle{When chatgpt meets smart contract vulnerability detection: How far are we?}
\newblock \bibinfo{journal}{\emph{arXiv preprint arXiv:2309.05520}} (\bibinfo{year}{2023}).
\newblock


\bibitem[Chen et~al\mbox{.}(2024)]%
        {chen2024identifying}
\bibfield{author}{\bibinfo{person}{Jiachi Chen}, \bibinfo{person}{Chong Chen}, \bibinfo{person}{Jiang Hu}, \bibinfo{person}{John Grundy}, \bibinfo{person}{Yanlin Wang}, \bibinfo{person}{Ting Chen}, {and} \bibinfo{person}{Zibin Zheng}.} \bibinfo{year}{2024}\natexlab{}.
\newblock \showarticletitle{Identifying Smart Contract Security Issues in Code Snippets from Stack Overflow}.
\newblock \bibinfo{journal}{\emph{arXiv preprint arXiv:2407.13271}} (\bibinfo{year}{2024}).
\newblock


\bibitem[Chen et~al\mbox{.}(2021)]%
        {chen2021evaluating}
\bibfield{author}{\bibinfo{person}{Mark Chen}, \bibinfo{person}{Jerry Tworek}, \bibinfo{person}{Heewoo Jun}, \bibinfo{person}{Qiming Yuan}, \bibinfo{person}{Henrique Ponde de~Oliveira Pinto}, \bibinfo{person}{Jared Kaplan}, \bibinfo{person}{Harri Edwards}, \bibinfo{person}{Yuri Burda}, \bibinfo{person}{Nicholas Joseph}, \bibinfo{person}{Greg Brockman}, {et~al\mbox{.}}} \bibinfo{year}{2021}\natexlab{}.
\newblock \showarticletitle{Evaluating large language models trained on code}.
\newblock \bibinfo{journal}{\emph{arXiv preprint arXiv:2107.03374}} (\bibinfo{year}{2021}).
\newblock


\bibitem[Cui et~al\mbox{.}(2024)]%
        {cui2024risk}
\bibfield{author}{\bibinfo{person}{Tianyu Cui}, \bibinfo{person}{Yanling Wang}, \bibinfo{person}{Chuanpu Fu}, \bibinfo{person}{Yong Xiao}, \bibinfo{person}{Sijia Li}, \bibinfo{person}{Xinhao Deng}, \bibinfo{person}{Yunpeng Liu}, \bibinfo{person}{Qinglin Zhang}, \bibinfo{person}{Ziyi Qiu}, \bibinfo{person}{Peiyang Li}, {et~al\mbox{.}}} \bibinfo{year}{2024}\natexlab{}.
\newblock \showarticletitle{Risk taxonomy, mitigation, and assessment benchmarks of large language model systems}.
\newblock \bibinfo{journal}{\emph{arXiv preprint arXiv:2401.05778}} (\bibinfo{year}{2024}).
\newblock


\bibitem[Dakhel et~al\mbox{.}(2023)]%
        {dakhel2023github}
\bibfield{author}{\bibinfo{person}{Arghavan~Moradi Dakhel}, \bibinfo{person}{Vahid Majdinasab}, \bibinfo{person}{Amin Nikanjam}, \bibinfo{person}{Foutse Khomh}, \bibinfo{person}{Michel~C Desmarais}, {and} \bibinfo{person}{Zhen Ming~Jack Jiang}.} \bibinfo{year}{2023}\natexlab{}.
\newblock \showarticletitle{Github copilot ai pair programmer: Asset or liability?}
\newblock \bibinfo{journal}{\emph{Journal of Systems and Software}}  \bibinfo{volume}{203} (\bibinfo{year}{2023}), \bibinfo{pages}{111734}.
\newblock


\bibitem[Daya~Guo(2024)]%
        {deepseek-coder}
\bibfield{author}{\bibinfo{person}{Dejian Yang Zhenda Xie Kai Dong Wentao Zhang Guanting Chen Xiao Bi Y. Wu Y.K. Li Fuli Luo Yingfei Xiong Wenfeng~Liang Daya~Guo, Qihao~Zhu}.} \bibinfo{year}{2024}\natexlab{}.
\newblock \bibinfo{title}{DeepSeek-Coder: When the Large Language Model Meets Programming -- The Rise of Code Intelligence}.
\newblock
\newblock
\urldef\tempurl%
\url{https://arxiv.org/abs/2401.14196}
\showURL{%
\tempurl}


\bibitem[Deng et~al\mbox{.}(2023)]%
        {deng2023jailbreaker}
\bibfield{author}{\bibinfo{person}{Gelei Deng}, \bibinfo{person}{Yi Liu}, \bibinfo{person}{Yuekang Li}, \bibinfo{person}{Kailong Wang}, \bibinfo{person}{Ying Zhang}, \bibinfo{person}{Zefeng Li}, \bibinfo{person}{Haoyu Wang}, \bibinfo{person}{Tianwei Zhang}, {and} \bibinfo{person}{Yang Liu}.} \bibinfo{year}{2023}\natexlab{}.
\newblock \showarticletitle{Jailbreaker: Automated jailbreak across multiple large language model chatbots}.
\newblock \bibinfo{journal}{\emph{arXiv preprint arXiv:2307.08715}} (\bibinfo{year}{2023}).
\newblock


\bibitem[Ding et~al\mbox{.}(2024)]%
        {ding2024crosscodeeval}
\bibfield{author}{\bibinfo{person}{Yangruibo Ding}, \bibinfo{person}{Zijian Wang}, \bibinfo{person}{Wasi Ahmad}, \bibinfo{person}{Hantian Ding}, \bibinfo{person}{Ming Tan}, \bibinfo{person}{Nihal Jain}, \bibinfo{person}{Murali~Krishna Ramanathan}, \bibinfo{person}{Ramesh Nallapati}, \bibinfo{person}{Parminder Bhatia}, \bibinfo{person}{Dan Roth}, {et~al\mbox{.}}} \bibinfo{year}{2024}\natexlab{}.
\newblock \showarticletitle{Crosscodeeval: A diverse and multilingual benchmark for cross-file code completion}.
\newblock \bibinfo{journal}{\emph{Advances in Neural Information Processing Systems}}  \bibinfo{volume}{36} (\bibinfo{year}{2024}).
\newblock


\bibitem[Du et~al\mbox{.}(2023)]%
        {du2023classeval}
\bibfield{author}{\bibinfo{person}{Xueying Du}, \bibinfo{person}{Mingwei Liu}, \bibinfo{person}{Kaixin Wang}, \bibinfo{person}{Hanlin Wang}, \bibinfo{person}{Junwei Liu}, \bibinfo{person}{Yixuan Chen}, \bibinfo{person}{Jiayi Feng}, \bibinfo{person}{Chaofeng Sha}, \bibinfo{person}{Xin Peng}, {and} \bibinfo{person}{Yiling Lou}.} \bibinfo{year}{2023}\natexlab{}.
\newblock \showarticletitle{Classeval: A manually-crafted benchmark for evaluating llms on class-level code generation}.
\newblock \bibinfo{journal}{\emph{arXiv preprint arXiv:2308.01861}} (\bibinfo{year}{2023}).
\newblock


\bibitem[EgoAlpha(2024)]%
        {GithubRepo1}
\bibfield{author}{\bibinfo{person}{GeshengSunDUT S A G A~R EgoAlpha, YFCao}.} \bibinfo{year}{2024}\natexlab{}.
\newblock \bibinfo{title}{prompt-in-context-learning}.
\newblock \bibinfo{howpublished}{\url{https://github.com/EgoAlpha/prompt-in-context-learning}}.
\newblock


\bibitem[Ethayarajh et~al\mbox{.}(2023)]%
        {ethayarajh2023stanford}
\bibfield{author}{\bibinfo{person}{Kawin Ethayarajh}, \bibinfo{person}{Heidi Zhang}, \bibinfo{person}{Yizhong Wang}, {and} \bibinfo{person}{Dan Jurafsky}.} \bibinfo{year}{2023}\natexlab{}.
\newblock \bibinfo{title}{Stanford human preferences dataset}.
\newblock
\newblock


\bibitem[Ganguli et~al\mbox{.}(2022)]%
        {ganguli2022red}
\bibfield{author}{\bibinfo{person}{Deep Ganguli}, \bibinfo{person}{Liane Lovitt}, \bibinfo{person}{Jackson Kernion}, \bibinfo{person}{Amanda Askell}, \bibinfo{person}{Yuntao Bai}, \bibinfo{person}{Saurav Kadavath}, \bibinfo{person}{Ben Mann}, \bibinfo{person}{Ethan Perez}, \bibinfo{person}{Nicholas Schiefer}, \bibinfo{person}{Kamal Ndousse}, {et~al\mbox{.}}} \bibinfo{year}{2022}\natexlab{}.
\newblock \showarticletitle{Red teaming language models to reduce harms: Methods, scaling behaviors, and lessons learned}.
\newblock \bibinfo{journal}{\emph{arXiv preprint arXiv:2209.07858}} (\bibinfo{year}{2022}).
\newblock


\bibitem[Grewal et~al\mbox{.}(2024)]%
        {grewal2024analyzing}
\bibfield{author}{\bibinfo{person}{Balreet Grewal}, \bibinfo{person}{Wentao Lu}, \bibinfo{person}{Sarah Nadi}, {and} \bibinfo{person}{Cor-Paul Bezemer}.} \bibinfo{year}{2024}\natexlab{}.
\newblock \showarticletitle{Analyzing Developer Use of ChatGPT Generated Code in Open Source GitHub Projects}.
\newblock  (\bibinfo{year}{2024}).
\newblock


\bibitem[Guo et~al\mbox{.}(2024)]%
        {guo2024exploring}
\bibfield{author}{\bibinfo{person}{Qi Guo}, \bibinfo{person}{Junming Cao}, \bibinfo{person}{Xiaofei Xie}, \bibinfo{person}{Shangqing Liu}, \bibinfo{person}{Xiaohong Li}, \bibinfo{person}{Bihuan Chen}, {and} \bibinfo{person}{Xin Peng}.} \bibinfo{year}{2024}\natexlab{}.
\newblock \showarticletitle{Exploring the potential of chatgpt in automated code refinement: An empirical study}. In \bibinfo{booktitle}{\emph{Proceedings of the 46th IEEE/ACM International Conference on Software Engineering}}. \bibinfo{pages}{1--13}.
\newblock


\bibitem[Ivison et~al\mbox{.}(2023)]%
        {ivison2023camels}
\bibfield{author}{\bibinfo{person}{Hamish Ivison}, \bibinfo{person}{Yizhong Wang}, \bibinfo{person}{Valentina Pyatkin}, \bibinfo{person}{Nathan Lambert}, \bibinfo{person}{Matthew Peters}, \bibinfo{person}{Pradeep Dasigi}, \bibinfo{person}{Joel Jang}, \bibinfo{person}{David Wadden}, \bibinfo{person}{Noah~A. Smith}, \bibinfo{person}{Iz Beltagy}, {and} \bibinfo{person}{Hannaneh Hajishirzi}.} \bibinfo{year}{2023}\natexlab{}.
\newblock \bibinfo{title}{Camels in a Changing Climate: Enhancing LM Adaptation with Tulu 2}.
\newblock
\newblock
\showeprint[arxiv]{2311.10702}~[cs.CL]


\bibitem[Ji et~al\mbox{.}(2024)]%
        {ji2024beavertails}
\bibfield{author}{\bibinfo{person}{Jiaming Ji}, \bibinfo{person}{Mickel Liu}, \bibinfo{person}{Josef Dai}, \bibinfo{person}{Xuehai Pan}, \bibinfo{person}{Chi Zhang}, \bibinfo{person}{Ce Bian}, \bibinfo{person}{Boyuan Chen}, \bibinfo{person}{Ruiyang Sun}, \bibinfo{person}{Yizhou Wang}, {and} \bibinfo{person}{Yaodong Yang}.} \bibinfo{year}{2024}\natexlab{}.
\newblock \showarticletitle{Beavertails: Towards improved safety alignment of llm via a human-preference dataset}.
\newblock \bibinfo{journal}{\emph{Advances in Neural Information Processing Systems}}  \bibinfo{volume}{36} (\bibinfo{year}{2024}).
\newblock


\bibitem[Lab(2024)]%
        {LLM-Safety-Leaderboard}
\bibfield{author}{\bibinfo{person}{Secure~Learning Lab}.} \bibinfo{year}{2024}\natexlab{}.
\newblock \bibinfo{title}{LLM Safety Leaderboard}.
\newblock
\newblock
\urldef\tempurl%
\url{https://huggingface.co/spaces/AI-Secure/llm-trustworthy-leaderboard}
\showURL{%
\tempurl}


\bibitem[Lai et~al\mbox{.}(2023)]%
        {lai2023ds}
\bibfield{author}{\bibinfo{person}{Yuhang Lai}, \bibinfo{person}{Chengxi Li}, \bibinfo{person}{Yiming Wang}, \bibinfo{person}{Tianyi Zhang}, \bibinfo{person}{Ruiqi Zhong}, \bibinfo{person}{Luke Zettlemoyer}, \bibinfo{person}{Wen-tau Yih}, \bibinfo{person}{Daniel Fried}, \bibinfo{person}{Sida Wang}, {and} \bibinfo{person}{Tao Yu}.} \bibinfo{year}{2023}\natexlab{}.
\newblock \showarticletitle{DS-1000: A natural and reliable benchmark for data science code generation}. In \bibinfo{booktitle}{\emph{International Conference on Machine Learning}}. PMLR, \bibinfo{pages}{18319--18345}.
\newblock


\bibitem[Li et~al\mbox{.}(2023)]%
        {li2023starcoder}
\bibfield{author}{\bibinfo{person}{Raymond Li}, \bibinfo{person}{Loubna~Ben Allal}, \bibinfo{person}{Yangtian Zi}, \bibinfo{person}{Niklas Muennighoff}, \bibinfo{person}{Denis Kocetkov}, \bibinfo{person}{Chenghao Mou}, \bibinfo{person}{Marc Marone}, \bibinfo{person}{Christopher Akiki}, \bibinfo{person}{Jia Li}, \bibinfo{person}{Jenny Chim}, {et~al\mbox{.}}} \bibinfo{year}{2023}\natexlab{}.
\newblock \showarticletitle{Starcoder: may the source be with you!}
\newblock \bibinfo{journal}{\emph{arXiv preprint arXiv:2305.06161}} (\bibinfo{year}{2023}).
\newblock


\bibitem[Lin et~al\mbox{.}(2023)]%
        {lin2023toxicchat}
\bibfield{author}{\bibinfo{person}{Zi Lin}, \bibinfo{person}{Zihan Wang}, \bibinfo{person}{Yongqi Tong}, \bibinfo{person}{Yangkun Wang}, \bibinfo{person}{Yuxin Guo}, \bibinfo{person}{Yujia Wang}, {and} \bibinfo{person}{Jingbo Shang}.} \bibinfo{year}{2023}\natexlab{}.
\newblock \bibinfo{title}{ToxicChat: Unveiling Hidden Challenges of Toxicity Detection in Real-World User-AI Conversation}.
\newblock
\newblock
\showeprint[arxiv]{2310.17389}~[cs.CL]


\bibitem[Liu et~al\mbox{.}(2020)]%
        {liu2020multi}
\bibfield{author}{\bibinfo{person}{Fang Liu}, \bibinfo{person}{Ge Li}, \bibinfo{person}{Yunfei Zhao}, {and} \bibinfo{person}{Zhi Jin}.} \bibinfo{year}{2020}\natexlab{}.
\newblock \showarticletitle{Multi-task learning based pre-trained language model for code completion}. In \bibinfo{booktitle}{\emph{Proceedings of the 35th IEEE/ACM International Conference on Automated Software Engineering}}. \bibinfo{pages}{473--485}.
\newblock


\bibitem[Liu et~al\mbox{.}(2024b)]%
        {liu2024lost}
\bibfield{author}{\bibinfo{person}{Nelson~F Liu}, \bibinfo{person}{Kevin Lin}, \bibinfo{person}{John Hewitt}, \bibinfo{person}{Ashwin Paranjape}, \bibinfo{person}{Michele Bevilacqua}, \bibinfo{person}{Fabio Petroni}, {and} \bibinfo{person}{Percy Liang}.} \bibinfo{year}{2024}\natexlab{b}.
\newblock \showarticletitle{Lost in the middle: How language models use long contexts}.
\newblock \bibinfo{journal}{\emph{Transactions of the Association for Computational Linguistics}}  \bibinfo{volume}{12} (\bibinfo{year}{2024}), \bibinfo{pages}{157--173}.
\newblock


\bibitem[Liu et~al\mbox{.}(2023b)]%
        {liu2023pre}
\bibfield{author}{\bibinfo{person}{Pengfei Liu}, \bibinfo{person}{Weizhe Yuan}, \bibinfo{person}{Jinlan Fu}, \bibinfo{person}{Zhengbao Jiang}, \bibinfo{person}{Hiroaki Hayashi}, {and} \bibinfo{person}{Graham Neubig}.} \bibinfo{year}{2023}\natexlab{b}.
\newblock \showarticletitle{Pre-train, prompt, and predict: A systematic survey of prompting methods in natural language processing}.
\newblock \bibinfo{journal}{\emph{Comput. Surveys}} \bibinfo{volume}{55}, \bibinfo{number}{9} (\bibinfo{year}{2023}), \bibinfo{pages}{1--35}.
\newblock


\bibitem[Liu et~al\mbox{.}(2024a)]%
        {liu2024empirical}
\bibfield{author}{\bibinfo{person}{Yongkun Liu}, \bibinfo{person}{Jiachi Chen}, \bibinfo{person}{Tingting Bi}, \bibinfo{person}{John Grundy}, \bibinfo{person}{Yanlin Wang}, \bibinfo{person}{Ting Chen}, \bibinfo{person}{Yutian Tang}, {and} \bibinfo{person}{Zibin Zheng}.} \bibinfo{year}{2024}\natexlab{a}.
\newblock \showarticletitle{An Empirical Study on Low Code Programming using Traditional vs Large Language Model Support}.
\newblock \bibinfo{journal}{\emph{arXiv preprint arXiv:2402.01156}} (\bibinfo{year}{2024}).
\newblock


\bibitem[Liu et~al\mbox{.}(2023a)]%
        {liu2023jailbreaking}
\bibfield{author}{\bibinfo{person}{Yi Liu}, \bibinfo{person}{Gelei Deng}, \bibinfo{person}{Zhengzi Xu}, \bibinfo{person}{Yuekang Li}, \bibinfo{person}{Yaowen Zheng}, \bibinfo{person}{Ying Zhang}, \bibinfo{person}{Lida Zhao}, \bibinfo{person}{Tianwei Zhang}, {and} \bibinfo{person}{Yang Liu}.} \bibinfo{year}{2023}\natexlab{a}.
\newblock \showarticletitle{Jailbreaking chatgpt via prompt engineering: An empirical study}.
\newblock \bibinfo{journal}{\emph{arXiv preprint arXiv:2305.13860}} (\bibinfo{year}{2023}).
\newblock


\bibitem[Lu et~al\mbox{.}(2021)]%
        {lu2021codexglue}
\bibfield{author}{\bibinfo{person}{Shuai Lu}, \bibinfo{person}{Daya Guo}, \bibinfo{person}{Shuo Ren}, \bibinfo{person}{Junjie Huang}, \bibinfo{person}{Alexey Svyatkovskiy}, \bibinfo{person}{Ambrosio Blanco}, \bibinfo{person}{Colin Clement}, \bibinfo{person}{Dawn Drain}, \bibinfo{person}{Daxin Jiang}, \bibinfo{person}{Duyu Tang}, {et~al\mbox{.}}} \bibinfo{year}{2021}\natexlab{}.
\newblock \showarticletitle{Codexglue: A machine learning benchmark dataset for code understanding and generation}.
\newblock \bibinfo{journal}{\emph{arXiv preprint arXiv:2102.04664}} (\bibinfo{year}{2021}).
\newblock


\bibitem[Mazeika et~al\mbox{.}(2024)]%
        {mazeika2024harmbench}
\bibfield{author}{\bibinfo{person}{Mantas Mazeika}, \bibinfo{person}{Long Phan}, \bibinfo{person}{Xuwang Yin}, \bibinfo{person}{Andy Zou}, \bibinfo{person}{Zifan Wang}, \bibinfo{person}{Norman Mu}, \bibinfo{person}{Elham Sakhaee}, \bibinfo{person}{Nathaniel Li}, \bibinfo{person}{Steven Basart}, \bibinfo{person}{Bo Li}, \bibinfo{person}{David Forsyth}, {and} \bibinfo{person}{Dan Hendrycks}.} \bibinfo{year}{2024}\natexlab{}.
\newblock \showarticletitle{HarmBench: A Standardized Evaluation Framework for Automated Red Teaming and Robust Refusal}.
\newblock  (\bibinfo{year}{2024}).
\newblock
\showeprint[arxiv]{2402.04249}~[cs.LG]


\bibitem[Microsoft(2024)]%
        {what-is-malware}
\bibfield{author}{\bibinfo{person}{Microsoft}.} \bibinfo{year}{2024}\natexlab{}.
\newblock \bibinfo{title}{what-is-malware}.
\newblock
\newblock
\urldef\tempurl%
\url{https://www.microsoft.com/zh-cn/security/business/security-101/what-is-malware}
\showURL{%
\tempurl}


\bibitem[Ning et~al\mbox{.}(2024)]%
        {ning2024mcgmarkencodablerobustonline}
\bibfield{author}{\bibinfo{person}{Kaiwen Ning}, \bibinfo{person}{Jiachi Chen}, \bibinfo{person}{Qingyuan Zhong}, \bibinfo{person}{Tao Zhang}, \bibinfo{person}{Yanlin Wang}, \bibinfo{person}{Wei Li}, \bibinfo{person}{Yu Zhang}, \bibinfo{person}{Weizhe Zhang}, {and} \bibinfo{person}{Zibin Zheng}.} \bibinfo{year}{2024}\natexlab{}.
\newblock \bibinfo{title}{MCGMark: An Encodable and Robust Online Watermark for LLM-Generated Malicious Code}.
\newblock
\newblock
\showeprint[arxiv]{2408.01354}~[cs.CR]
\urldef\tempurl%
\url{https://arxiv.org/abs/2408.01354}
\showURL{%
\tempurl}


\bibitem[OpenAI(2024)]%
        {openai-api-interface}
\bibfield{author}{\bibinfo{person}{OpenAI}.} \bibinfo{year}{2024}\natexlab{}.
\newblock \showarticletitle{Openai api interface}.
\newblock  (\bibinfo{year}{2024}).
\newblock
\urldef\tempurl%
\url{https://platform.openai.com/docs/api-reference}
\showURL{%
\tempurl}


\bibitem[Ostendorf et~al\mbox{.}(2005)]%
        {ostendorf2005human}
\bibfield{author}{\bibinfo{person}{Mari Ostendorf}, \bibinfo{person}{Elizabeth Shriberg}, {and} \bibinfo{person}{Andreas Stolcke}.} \bibinfo{year}{2005}\natexlab{}.
\newblock \showarticletitle{Human language technology: Opportunities and challenges}. In \bibinfo{booktitle}{\emph{Proceedings.(ICASSP'05). IEEE International Conference on Acoustics, Speech, and Signal Processing, 2005.}}, Vol.~\bibinfo{volume}{5}. IEEE, \bibinfo{pages}{v--949}.
\newblock


\bibitem[Ouyang et~al\mbox{.}(2022)]%
        {ouyang2022training}
\bibfield{author}{\bibinfo{person}{Long Ouyang}, \bibinfo{person}{Jeffrey Wu}, \bibinfo{person}{Xu Jiang}, \bibinfo{person}{Diogo Almeida}, \bibinfo{person}{Carroll Wainwright}, \bibinfo{person}{Pamela Mishkin}, \bibinfo{person}{Chong Zhang}, \bibinfo{person}{Sandhini Agarwal}, \bibinfo{person}{Katarina Slama}, \bibinfo{person}{Alex Ray}, {et~al\mbox{.}}} \bibinfo{year}{2022}\natexlab{}.
\newblock \showarticletitle{Training language models to follow instructions with human feedback}.
\newblock \bibinfo{journal}{\emph{Advances in neural information processing systems}}  \bibinfo{volume}{35} (\bibinfo{year}{2022}), \bibinfo{pages}{27730--27744}.
\newblock


\bibitem[Parrish et~al\mbox{.}(2021)]%
        {parrish2021bbq}
\bibfield{author}{\bibinfo{person}{Alicia Parrish}, \bibinfo{person}{Angelica Chen}, \bibinfo{person}{Nikita Nangia}, \bibinfo{person}{Vishakh Padmakumar}, \bibinfo{person}{Jason Phang}, \bibinfo{person}{Jana Thompson}, \bibinfo{person}{Phu~Mon Htut}, {and} \bibinfo{person}{Samuel~R Bowman}.} \bibinfo{year}{2021}\natexlab{}.
\newblock \showarticletitle{BBQ: A hand-built bias benchmark for question answering}.
\newblock \bibinfo{journal}{\emph{arXiv preprint arXiv:2110.08193}} (\bibinfo{year}{2021}).
\newblock


\bibitem[Pearce et~al\mbox{.}(2023)]%
        {pearce2023examining}
\bibfield{author}{\bibinfo{person}{Hammond Pearce}, \bibinfo{person}{Benjamin Tan}, \bibinfo{person}{Baleegh Ahmad}, \bibinfo{person}{Ramesh Karri}, {and} \bibinfo{person}{Brendan Dolan-Gavitt}.} \bibinfo{year}{2023}\natexlab{}.
\newblock \showarticletitle{Examining zero-shot vulnerability repair with large language models}. In \bibinfo{booktitle}{\emph{2023 IEEE Symposium on Security and Privacy (SP)}}. IEEE, \bibinfo{pages}{2339--2356}.
\newblock


\bibitem[Puttaparthi et~al\mbox{.}(2023)]%
        {puttaparthi2023comprehensive}
\bibfield{author}{\bibinfo{person}{Poorna Chander~Reddy Puttaparthi}, \bibinfo{person}{Soham~Sanjay Deo}, \bibinfo{person}{Hakan Gul}, \bibinfo{person}{Yiming Tang}, \bibinfo{person}{Weiyi Shang}, {and} \bibinfo{person}{Zhe Yu}.} \bibinfo{year}{2023}\natexlab{}.
\newblock \showarticletitle{Comprehensive evaluation of chatgpt reliability through multilingual inquiries}.
\newblock \bibinfo{journal}{\emph{arXiv preprint arXiv:2312.10524}} (\bibinfo{year}{2023}).
\newblock


\bibitem[Rafailov et~al\mbox{.}(2024)]%
        {rafailov2024direct}
\bibfield{author}{\bibinfo{person}{Rafael Rafailov}, \bibinfo{person}{Archit Sharma}, \bibinfo{person}{Eric Mitchell}, \bibinfo{person}{Christopher~D Manning}, \bibinfo{person}{Stefano Ermon}, {and} \bibinfo{person}{Chelsea Finn}.} \bibinfo{year}{2024}\natexlab{}.
\newblock \showarticletitle{Direct preference optimization: Your language model is secretly a reward model}.
\newblock \bibinfo{journal}{\emph{Advances in Neural Information Processing Systems}}  \bibinfo{volume}{36} (\bibinfo{year}{2024}).
\newblock


\bibitem[R{\"o}ttger et~al\mbox{.}(2024)]%
        {rottger2024safetyprompts}
\bibfield{author}{\bibinfo{person}{Paul R{\"o}ttger}, \bibinfo{person}{Fabio Pernisi}, \bibinfo{person}{Bertie Vidgen}, {and} \bibinfo{person}{Dirk Hovy}.} \bibinfo{year}{2024}\natexlab{}.
\newblock \showarticletitle{SafetyPrompts: a Systematic Review of Open Datasets for Evaluating and Improving Large Language Model Safety}.
\newblock \bibinfo{journal}{\emph{arXiv preprint arXiv:2404.05399}} (\bibinfo{year}{2024}).
\newblock


\bibitem[Roziere et~al\mbox{.}(2023)]%
        {roziere2023code}
\bibfield{author}{\bibinfo{person}{Baptiste Roziere}, \bibinfo{person}{Jonas Gehring}, \bibinfo{person}{Fabian Gloeckle}, \bibinfo{person}{Sten Sootla}, \bibinfo{person}{Itai Gat}, \bibinfo{person}{Xiaoqing~Ellen Tan}, \bibinfo{person}{Yossi Adi}, \bibinfo{person}{Jingyu Liu}, \bibinfo{person}{Tal Remez}, \bibinfo{person}{J{\'e}r{\'e}my Rapin}, {et~al\mbox{.}}} \bibinfo{year}{2023}\natexlab{}.
\newblock \showarticletitle{Code llama: Open foundation models for code}.
\newblock \bibinfo{journal}{\emph{arXiv preprint arXiv:2308.12950}} (\bibinfo{year}{2023}).
\newblock


\bibitem[Scherrer et~al\mbox{.}(2023)]%
        {scherrer2023moralchoice}
\bibfield{author}{\bibinfo{person}{Nino Scherrer}, \bibinfo{person}{Claudia Shi}, \bibinfo{person}{Amir Feder}, {and} \bibinfo{person}{David Blei}.} \bibinfo{year}{2023}\natexlab{}.
\newblock \bibinfo{title}{Evaluating the Moral Beliefs Encoded in LLMs}.
\newblock
\newblock


\bibitem[Shen et~al\mbox{.}(2023)]%
        {shen2023large}
\bibfield{author}{\bibinfo{person}{Tianhao Shen}, \bibinfo{person}{Renren Jin}, \bibinfo{person}{Yufei Huang}, \bibinfo{person}{Chuang Liu}, \bibinfo{person}{Weilong Dong}, \bibinfo{person}{Zishan Guo}, \bibinfo{person}{Xinwei Wu}, \bibinfo{person}{Yan Liu}, {and} \bibinfo{person}{Deyi Xiong}.} \bibinfo{year}{2023}\natexlab{}.
\newblock \showarticletitle{Large language model alignment: A survey}.
\newblock \bibinfo{journal}{\emph{arXiv preprint arXiv:2309.15025}} (\bibinfo{year}{2023}).
\newblock


\bibitem[Team(2023)]%
        {MosaicML2023Introducing}
\bibfield{author}{\bibinfo{person}{MosaicML~NLP Team}.} \bibinfo{year}{2023}\natexlab{}.
\newblock \bibinfo{booktitle}{\emph{Introducing MPT-7B: A New Standard for Open-Source, Commercially Usable LLMs}}.
\newblock
\urldef\tempurl%
\url{www.mosaicml.com/blog/mpt-7b}
\showURL{%
\tempurl}
\newblock
\shownote{Accessed: 2023-03-28}.


\bibitem[Touvron et~al\mbox{.}(2023)]%
        {touvron2023llama}
\bibfield{author}{\bibinfo{person}{Hugo Touvron}, \bibinfo{person}{Louis Martin}, \bibinfo{person}{Kevin Stone}, \bibinfo{person}{Peter Albert}, \bibinfo{person}{Amjad Almahairi}, \bibinfo{person}{Yasmine Babaei}, \bibinfo{person}{Nikolay Bashlykov}, \bibinfo{person}{Soumya Batra}, \bibinfo{person}{Prajjwal Bhargava}, \bibinfo{person}{Shruti Bhosale}, {et~al\mbox{.}}} \bibinfo{year}{2023}\natexlab{}.
\newblock \showarticletitle{Llama 2: Open foundation and fine-tuned chat models}.
\newblock \bibinfo{journal}{\emph{arXiv preprint arXiv:2307.09288}} (\bibinfo{year}{2023}).
\newblock


\bibitem[Tunstall et~al\mbox{.}(2023)]%
        {tunstall2023zephyr}
\bibfield{author}{\bibinfo{person}{Lewis Tunstall}, \bibinfo{person}{Edward Beeching}, \bibinfo{person}{Nathan Lambert}, \bibinfo{person}{Nazneen Rajani}, \bibinfo{person}{Kashif Rasul}, \bibinfo{person}{Younes Belkada}, \bibinfo{person}{Shengyi Huang}, \bibinfo{person}{Leandro von Werra}, \bibinfo{person}{Clémentine Fourrier}, \bibinfo{person}{Nathan Habib}, \bibinfo{person}{Nathan Sarrazin}, \bibinfo{person}{Omar Sanseviero}, \bibinfo{person}{Alexander~M. Rush}, {and} \bibinfo{person}{Thomas Wolf}.} \bibinfo{year}{2023}\natexlab{}.
\newblock \bibinfo{title}{Zephyr: Direct Distillation of LM Alignment}.
\newblock
\newblock
\showeprint[arxiv]{2310.16944}~[cs.LG]


\bibitem[Vaswani et~al\mbox{.}(2017)]%
        {vaswani2017attention}
\bibfield{author}{\bibinfo{person}{Ashish Vaswani}, \bibinfo{person}{Noam Shazeer}, \bibinfo{person}{Niki Parmar}, \bibinfo{person}{Jakob Uszkoreit}, \bibinfo{person}{Llion Jones}, \bibinfo{person}{Aidan~N Gomez}, \bibinfo{person}{{\L}ukasz Kaiser}, {and} \bibinfo{person}{Illia Polosukhin}.} \bibinfo{year}{2017}\natexlab{}.
\newblock \showarticletitle{Attention is all you need}.
\newblock \bibinfo{journal}{\emph{Advances in neural information processing systems}}  \bibinfo{volume}{30} (\bibinfo{year}{2017}).
\newblock


\bibitem[Wang et~al\mbox{.}(2023)]%
        {wang2023decodingtrust}
\bibfield{author}{\bibinfo{person}{Boxin Wang}, \bibinfo{person}{Weixin Chen}, \bibinfo{person}{Hengzhi Pei}, \bibinfo{person}{Chulin Xie}, \bibinfo{person}{Mintong Kang}, \bibinfo{person}{Chenhui Zhang}, \bibinfo{person}{Chejian Xu}, \bibinfo{person}{Zidi Xiong}, \bibinfo{person}{Ritik Dutta}, \bibinfo{person}{Rylan Schaeffer}, {et~al\mbox{.}}} \bibinfo{year}{2023}\natexlab{}.
\newblock \showarticletitle{Decodingtrust: A comprehensive assessment of trustworthiness in gpt models}.
\newblock \bibinfo{journal}{\emph{arXiv preprint arXiv:2306.11698}} (\bibinfo{year}{2023}).
\newblock


\bibitem[Wang et~al\mbox{.}(2024)]%
        {wang2024rlcoderreinforcementlearningrepositorylevel}
\bibfield{author}{\bibinfo{person}{Yanlin Wang}, \bibinfo{person}{Yanli Wang}, \bibinfo{person}{Daya Guo}, \bibinfo{person}{Jiachi Chen}, \bibinfo{person}{Ruikai Zhang}, \bibinfo{person}{Yuchi Ma}, {and} \bibinfo{person}{Zibin Zheng}.} \bibinfo{year}{2024}\natexlab{}.
\newblock \bibinfo{title}{RLCoder: Reinforcement Learning for Repository-Level Code Completion}.
\newblock
\newblock
\showeprint[arxiv]{2407.19487}~[cs.SE]
\urldef\tempurl%
\url{https://arxiv.org/abs/2407.19487}
\showURL{%
\tempurl}


\bibitem[Wei et~al\mbox{.}(2024)]%
        {wei2024jailbroken}
\bibfield{author}{\bibinfo{person}{Alexander Wei}, \bibinfo{person}{Nika Haghtalab}, {and} \bibinfo{person}{Jacob Steinhardt}.} \bibinfo{year}{2024}\natexlab{}.
\newblock \showarticletitle{Jailbroken: How does llm safety training fail?}
\newblock \bibinfo{journal}{\emph{Advances in Neural Information Processing Systems}}  \bibinfo{volume}{36} (\bibinfo{year}{2024}).
\newblock


\bibitem[Wei et~al\mbox{.}(2022)]%
        {wei2022chain}
\bibfield{author}{\bibinfo{person}{Jason Wei}, \bibinfo{person}{Xuezhi Wang}, \bibinfo{person}{Dale Schuurmans}, \bibinfo{person}{Maarten Bosma}, \bibinfo{person}{Fei Xia}, \bibinfo{person}{Ed Chi}, \bibinfo{person}{Quoc~V Le}, \bibinfo{person}{Denny Zhou}, {et~al\mbox{.}}} \bibinfo{year}{2022}\natexlab{}.
\newblock \showarticletitle{Chain-of-thought prompting elicits reasoning in large language models}.
\newblock \bibinfo{journal}{\emph{Advances in neural information processing systems}}  \bibinfo{volume}{35} (\bibinfo{year}{2022}), \bibinfo{pages}{24824--24837}.
\newblock


\bibitem[Yang et~al\mbox{.}(2024)]%
        {yang2024hyperionunveilingdappinconsistencies}
\bibfield{author}{\bibinfo{person}{Shuo Yang}, \bibinfo{person}{Xingwei Lin}, \bibinfo{person}{Jiachi Chen}, \bibinfo{person}{Qingyuan Zhong}, \bibinfo{person}{Lei Xiao}, \bibinfo{person}{Renke Huang}, \bibinfo{person}{Yanlin Wang}, {and} \bibinfo{person}{Zibin Zheng}.} \bibinfo{year}{2024}\natexlab{}.
\newblock \bibinfo{title}{Hyperion: Unveiling DApp Inconsistencies using LLM and Dataflow-Guided Symbolic Execution}.
\newblock
\newblock
\showeprint[arxiv]{2408.06037}~[cs.SE]
\urldef\tempurl%
\url{https://arxiv.org/abs/2408.06037}
\showURL{%
\tempurl}


\bibitem[Yu et~al\mbox{.}(2024)]%
        {yu2024codereval}
\bibfield{author}{\bibinfo{person}{Hao Yu}, \bibinfo{person}{Bo Shen}, \bibinfo{person}{Dezhi Ran}, \bibinfo{person}{Jiaxin Zhang}, \bibinfo{person}{Qi Zhang}, \bibinfo{person}{Yuchi Ma}, \bibinfo{person}{Guangtai Liang}, \bibinfo{person}{Ying Li}, \bibinfo{person}{Qianxiang Wang}, {and} \bibinfo{person}{Tao Xie}.} \bibinfo{year}{2024}\natexlab{}.
\newblock \showarticletitle{Codereval: A benchmark of pragmatic code generation with generative pre-trained models}. In \bibinfo{booktitle}{\emph{Proceedings of the 46th IEEE/ACM International Conference on Software Engineering}}. \bibinfo{pages}{1--12}.
\newblock


\bibitem[Yu et~al\mbox{.}(2023)]%
        {yu2023gptfuzzer}
\bibfield{author}{\bibinfo{person}{Jiahao Yu}, \bibinfo{person}{Xingwei Lin}, {and} \bibinfo{person}{Xinyu Xing}.} \bibinfo{year}{2023}\natexlab{}.
\newblock \showarticletitle{Gptfuzzer: Red teaming large language models with auto-generated jailbreak prompts}.
\newblock \bibinfo{journal}{\emph{arXiv preprint arXiv:2309.10253}} (\bibinfo{year}{2023}).
\newblock


\bibitem[Zhang et~al\mbox{.}(2023a)]%
        {zhang2023repocoder}
\bibfield{author}{\bibinfo{person}{Fengji Zhang}, \bibinfo{person}{Bei Chen}, \bibinfo{person}{Yue Zhang}, \bibinfo{person}{Jacky Keung}, \bibinfo{person}{Jin Liu}, \bibinfo{person}{Daoguang Zan}, \bibinfo{person}{Yi Mao}, \bibinfo{person}{Jian-Guang Lou}, {and} \bibinfo{person}{Weizhu Chen}.} \bibinfo{year}{2023}\natexlab{a}.
\newblock \showarticletitle{Repocoder: Repository-level code completion through iterative retrieval and generation}.
\newblock \bibinfo{journal}{\emph{arXiv preprint arXiv:2303.12570}} (\bibinfo{year}{2023}).
\newblock


\bibitem[Zhang et~al\mbox{.}(2023b)]%
        {zhang2023safetybench}
\bibfield{author}{\bibinfo{person}{Zhexin Zhang}, \bibinfo{person}{Leqi Lei}, \bibinfo{person}{Lindong Wu}, \bibinfo{person}{Rui Sun}, \bibinfo{person}{Yongkang Huang}, \bibinfo{person}{Chong Long}, \bibinfo{person}{Xiao Liu}, \bibinfo{person}{Xuanyu Lei}, \bibinfo{person}{Jie Tang}, {and} \bibinfo{person}{Minlie Huang}.} \bibinfo{year}{2023}\natexlab{b}.
\newblock \showarticletitle{Safetybench: Evaluating the safety of large language models with multiple choice questions}.
\newblock \bibinfo{journal}{\emph{arXiv preprint arXiv:2309.07045}} (\bibinfo{year}{2023}).
\newblock


\bibitem[Zhao et~al\mbox{.}(2023)]%
        {LLMSurvey}
\bibfield{author}{\bibinfo{person}{Wayne~Xin Zhao}, \bibinfo{person}{Kun Zhou}, \bibinfo{person}{Junyi Li}, \bibinfo{person}{Tianyi Tang}, \bibinfo{person}{Xiaolei Wang}, \bibinfo{person}{Yupeng Hou}, \bibinfo{person}{Yingqian Min}, \bibinfo{person}{Beichen Zhang}, \bibinfo{person}{Junjie Zhang}, \bibinfo{person}{Zican Dong}, \bibinfo{person}{Yifan Du}, \bibinfo{person}{Chen Yang}, \bibinfo{person}{Yushuo Chen}, \bibinfo{person}{Zhipeng Chen}, \bibinfo{person}{Jinhao Jiang}, \bibinfo{person}{Ruiyang Ren}, \bibinfo{person}{Yifan Li}, \bibinfo{person}{Xinyu Tang}, \bibinfo{person}{Zikang Liu}, \bibinfo{person}{Peiyu Liu}, \bibinfo{person}{Jian-Yun Nie}, {and} \bibinfo{person}{Ji-Rong Wen}.} \bibinfo{year}{2023}\natexlab{}.
\newblock \showarticletitle{A Survey of Large Language Models}.
\newblock \bibinfo{journal}{\emph{arXiv preprint arXiv:2303.18223}} (\bibinfo{year}{2023}).
\newblock
\urldef\tempurl%
\url{http://arxiv.org/abs/2303.18223}
\showURL{%
\tempurl}


\bibitem[Zheng et~al\mbox{.}(2024)]%
        {zheng2024judging}
\bibfield{author}{\bibinfo{person}{Lianmin Zheng}, \bibinfo{person}{Wei-Lin Chiang}, \bibinfo{person}{Ying Sheng}, \bibinfo{person}{Siyuan Zhuang}, \bibinfo{person}{Zhanghao Wu}, \bibinfo{person}{Yonghao Zhuang}, \bibinfo{person}{Zi Lin}, \bibinfo{person}{Zhuohan Li}, \bibinfo{person}{Dacheng Li}, \bibinfo{person}{Eric Xing}, {et~al\mbox{.}}} \bibinfo{year}{2024}\natexlab{}.
\newblock \showarticletitle{Judging llm-as-a-judge with mt-bench and chatbot arena}.
\newblock \bibinfo{journal}{\emph{Advances in Neural Information Processing Systems}}  \bibinfo{volume}{36} (\bibinfo{year}{2024}).
\newblock


\bibitem[Zheng et~al\mbox{.}(2023a)]%
        {zheng2023towards}
\bibfield{author}{\bibinfo{person}{Zibin Zheng}, \bibinfo{person}{Kaiwen Ning}, \bibinfo{person}{Jiachi Chen}, \bibinfo{person}{Yanlin Wang}, \bibinfo{person}{Wenqing Chen}, \bibinfo{person}{Lianghong Guo}, {and} \bibinfo{person}{Weicheng Wang}.} \bibinfo{year}{2023}\natexlab{a}.
\newblock \showarticletitle{Towards an understanding of large language models in software engineering tasks}.
\newblock \bibinfo{journal}{\emph{arXiv preprint arXiv:2308.11396}} (\bibinfo{year}{2023}).
\newblock


\bibitem[Zheng et~al\mbox{.}(2023b)]%
        {zheng2023survey}
\bibfield{author}{\bibinfo{person}{Zibin Zheng}, \bibinfo{person}{Kaiwen Ning}, \bibinfo{person}{Yanlin Wang}, \bibinfo{person}{Jingwen Zhang}, \bibinfo{person}{Dewu Zheng}, \bibinfo{person}{Mingxi Ye}, {and} \bibinfo{person}{Jiachi Chen}.} \bibinfo{year}{2023}\natexlab{b}.
\newblock \showarticletitle{A survey of large language models for code: Evolution, benchmarking, and future trends}.
\newblock \bibinfo{journal}{\emph{arXiv preprint arXiv:2311.10372}} (\bibinfo{year}{2023}).
\newblock


\bibitem[Zhuo(2024)]%
        {zhuo2024ice}
\bibfield{author}{\bibinfo{person}{Terry~Yue Zhuo}.} \bibinfo{year}{2024}\natexlab{}.
\newblock \showarticletitle{ICE-Score: Instructing Large Language Models to Evaluate Code}. In \bibinfo{booktitle}{\emph{Findings of the Association for Computational Linguistics: EACL 2024}}. \bibinfo{pages}{2232--2242}.
\newblock


\end{thebibliography}
\bibliographystyle{ACM-Reference-Format}


\end{CJK}
\end{document}